\documentclass[debug]{rmaa}

\usepackage{longtable,lscape}
\usepackage{multirow}

\title{Planetary nebulae in the inner Milky Way: new abundances}
\author{
  O. Cavichia,\altaffilmark{1} 
  R. D. D. Costa,\altaffilmark{1} 
  and W. J. Maciel\altaffilmark{1}}

\altaffiltext{1}{Instituto de Astronomia, Geof\'\i sica e Ci\^encias
                 Atmosf\'ericas, Universidade de S\~ao Paulo, Brazil.}

\shortauthor{Cavichia, Costa \& Maciel}
\shorttitle{Planetary nebulae in the inner Milky Way}

\fulladdresses{
\item Departamento de Astronomia, Instituto de Astronomia, Geof\'isica e Ci\^encias Atmosf\'ericas, Universidade de S\~ao Paulo, Rua do Mat\~ao, 1226, Cidade Universit\'aria, 05508-900 S\~ao Paulo-SP, Brazil.
}

\listofauthors{O. Cavichia, R. D. D. Costa, \& W. J. Maciel}
\indexauthor{Cavichia, O.}
\indexauthor{Costa, R. D. D.}
\indexauthor{Maciel, W. J.}

\abstract{The study of planetary nebulae in the inner-disk and bulge gives important information on the chemical abundances of elements such as He, N, O, Ar, Ne, and on the evolution of these abundances, which is associated with the evolution of intermediate-mass stars and the chemical evolution of the Galaxy.

 We present accurate abundances of the elements He, N, S, O, Ar, and Ne for a sample of 54 planetary nebulae located towards the bulge of the Galaxy, for which 33 have the abundances derived for the first time. The abundances are derived based on observations in the optical domain made at the National Laboratory for Astrophysics (LNA, Brazil). The data show a good agreement with other results in the literature, in the sense that the distribution of the abundances is similar to those works. 
}

\resumen{}

\addkeyword{Galaxy: abundances}
\addkeyword{Galaxy: evolution}
\addkeyword{planetary nebulae: general}
\addkeyword{techniques: spectroscopic}

\begin{document}
\maketitle

\section{Introduction}
\label{sec:intro}

The bulge of the Galaxy shows a large metallicity dispersion. The study of the metallicity distribution from K giants, as done by Rich~(\citeyear{rich88}), shows values from 0.1 to 10 Z$_{\odot}$. More recently, Rich \& Origlia~(\citeyear{rich05})  find an $\alpha$-enhancement at the level of $+0.3$ dex relative to the solar composition stars for 14 M giants and within a narrow metallicity range around [Fe/H$]=-0.2$.   
Zoccali et al.~(2006) and Lecureur et al.~(\citeyear{lecu07}) find that bulge stars have larger values of [O/Fe] and [Mg/Fe] when compared to thin and thick disk stars. This is the signature of a chemical enrichment by massive stars, progenitors of type II supernovae, with little or no contribution from type Ia supernovae, showing a shorter formation timescale for the bulge than both thin and thick disks. 

In this context, planetary nebulae (PNe) are an important tool for the study of the chemical evolution of galaxies. The understanding of this stage of stellar evolution allows us to grasp how the Galaxy originated and developed. As an intermediate mass star evolution product, PNe offer the possibility of studying both elements produced in low and intermediate mass stars, such as helium and nitrogen, and also elements which result from the nucleosynthesis of large mass stars, such as oxygen, sulfur and neon, which are present in the interstellar medium at formation epoch of the PNe stellar progenitor. 

Regardless of the fact that the chemical abundances obtained from PNe are relatively accurate, their distances are subject of discussion even nowadays. Excluding a few PNe whose distances are determined from direct methods such as trigonometric parallax or in cases where there is a binary companion in the main sequence, most PNe have their distances derived from nebular properties (see e.g. Maciel \& Pottasch~\citeyear{maciel80}, Cahn et al.~\citeyear{cahn92}, Stanghellini et al.~\citeyear{stangh08}). These uncertainties in the distances of PNe make the study of the chemical properties with respect to the galactocentric distance a difficult task. In spite of the uncertainties, statistical distance scales are still the best tool to study the chemical abundance patterns in the Galaxy from the point of view of PNe, as e.g. done by Maciel \& Quireza~(\citeyear{maciel99}), Maciel et al.~(\citeyear{maciel06}), Perinotto \& Morbidelli~(\citeyear{peri06}), and Gutenkunst et al.~(\citeyear{guten08}).

Since the bulge and the disk may have different evolution histories, described for example by the disk inside out formation model (Chiappini et al.~\citeyear{chiap01}) or by the multiple infalls scenario (Costa et al.~\citeyear{costa05},\citeyear{costa08}), we should expect these differences reflected on the chemical properties of each component. Indeed, bulge and disk display different chemical abundance patterns like the radial abundance gradients found in the disk (Carigi et al. \citeyear{carigi05}; Daflon \& Cunha~\citeyear{daflon04}; Andrievsky et al.~\citeyear{andri04}; Maciel et al.~\citeyear{maciel05},\citeyear{maciel06}), or the large abundance distribution found in the bulge (Rich~\citeyear{rich88}, Zoccali et al.~\citeyear{zoc03,zoc06}). 

On the other hand, Chiappini et al.~(\citeyear{chiap09}) made a comparison between abundances from PNe located at the bulge, inner-disk and Large Magellanic Cloud. Their results do not show any clear difference between bulge and inner-disk objects. Some other previous studies of the Galactic bulge based on abundances of PNe such as Ratag et al.~(\citeyear{ratag92}), Cuisinier et al.~(\citeyear{cuisin00}), Escudero \& Costa (\citeyear{escu01}), Escudero et al. (\citeyear{escu04}), and Exter et al. (\citeyear{exter04}), find that bulge PNe have an abundance distribution similar to disk PNe, showing that He, O, Si, Ar, and Ca have a normal abundance pattern, favouring therefore a slower Galactic evolution than that indicated by stars. In conclusion, the study of chemical abundances in the inner region of the Galaxy is still an open question, especially regarding the bulge-disk connection. 

The goal of this paper is to report new spectrophotometric observations for a sample of PNe located in the inner-disk and bulge of the Milky Way Galaxy, aiming to derive their nebular physical parameters and chemical abundances, as has been done by our group (see e.g. Costa et al. \citeyear{costa96,costa00}, Escudero \& Costa \citeyear{escu01}, Escudero et al.~\citeyear{escu04}, and references therein), as part of a long-term program to derive a large sample of chemical abundances of southern PNe. As a result, our database has become one of the largest in the literature with a very homogeneous observational setup, reduction and analysis procedures, which is necessary to perform large scale statistical studies. In this work, 33 objects have their abundances derived for the first time. Additionally, objects in common with other samples are used to compare our data with previously data already published. The comparison of the final abundances with those obtained in other multi-object studies allowed us to assess the accuracy of the new abundances.

This paper is organized as follows: in \S~\ref{sec:obs_red} the details of the observations and data reduction procedures are presented. In \S~3 we describe the process of determination of chemical abundances and the new abundances are listed. In \S~4, a comparison is made between the abundances obtained in this work and those taken from the literature. Finally, in \S~5 the main conclusions are presented.

\section{Observations and data reduction}
\label{sec:obs_red}

\subsection{Observations}
\label{sec:obs}

The observations were made at the 1.60 m telescope of the National Laboratory for Astrophysics (LNA, Brazil) during 2006 and 2007, according to the log of observations shown by table \ref{tab_log}. In this table, column 1 displays the PN G designation, column 2 the usual name, columns 3 and 4 the equatorial coordinates for epoch 2000, column 5 the date of observation, and column 6 the exposure time in seconds. A Cassegrain Boller \& Chivens spectrograph was used with a 300 l/mm grid, which provides a reciprocal dispersion of 0.2 nm/pixel. For all program objects, a long slit of 1.5 arcsec width was used. Each night at least three spectrophotometric standard stars were observed to improve the flux calibration. These stars were observed with a long slit of 7.5 arcsec width, allowing a more precise flux calibration.

\addtocounter{table}{1}

The sample was selected from the \emph{Strasbourg - ESO catalogue of galactic planetary nebulae} (Acker et al.~\citeyear{acker92}), based on three criteria: galactic coordinates within the range $|\ell| \leq 25^{\circ}$ and $|b| \leq 10^{\circ}$, 5 GHz flux below 100 mJy, and optical diameter lower than 12 arcsec. The galactic coordinates were used to take into account only the PNe which are in the galactic center direction. The combination of the other two criteria leads to the rejection of about 90--95\% of the PNe which are in the galactic center direction, but have heliocentric distances lower than 4 kpc (cf. Stasi\'nska et al.~\citeyear{stas98}). These criteria are commonly used by other authors to select bulge PNe (e.g. Exter et al.~\citeyear{exter04} and Chiappini et al.~\citeyear{chiap09}). Hence most of the objects selected in this work should be at or near the bulge.

\begin{figure}[!ht]
   \centering
   \includegraphics[width=9cm]{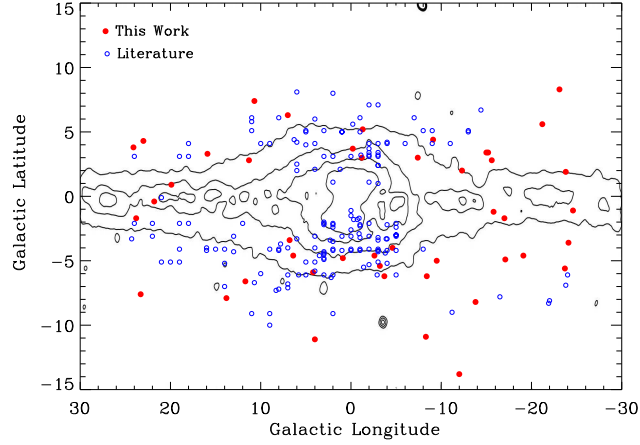}
   \caption{PNe distribution with respect to the galactic bulge for this work (filled circles) and data from the literature (open circles). }
    \label{pne_coord}%
\end{figure}

Figure \ref{pne_coord} displays the distribution of the sample with respect to the galactic bulge. The figure also shows the distribution for the objects selected from the literature (see \S~\ref{discussion_data} for more details). As can be seen, the selected objects are in the direction of the galactic bulge, whose contours are displayed using the image from the 2.2 $\mu$m COBE/DIRBE satellite  plot (Weiland et al.~\citeyear{wei94}). Furthermore, the sample objects spread over the entire region of the galactic bulge, avoiding tendencies in the chemical abundances analysis introduced by partial coverage of the bulge, as found by Escudero \& Costa (\citeyear{escu01}). They showed that objects located in a region with galactic latitude larger than 5 degrees display lower abundances  when compared with other works in the literature such as Ratag et al.~(\citeyear{ratag97}), Cuisinier et al.~(\citeyear{cuisin00}), and Stasi{\'n}ska et al.~(\citeyear{stas98}), whose samples were located elsewhere. 

\addtocounter{table}{1}

\begin{figure}[!ht]
   \centering
   \includegraphics[width=9cm]{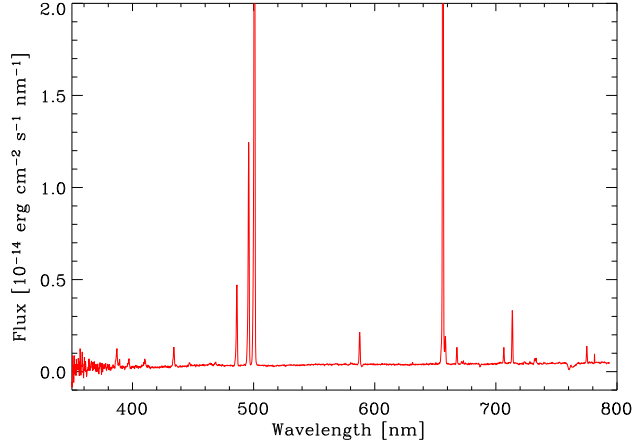}
   \caption{Typical spectrum from our sample for the object Pe 1-15. [\ion{O}{iii}] 500.7 nm and H$\alpha$ lines are saturated for a better view.}
    \label{pe115spec}%
\end{figure}

Data reduction was performed using the IRAF package, following the standard procedure for long slit spectra: correction of bias, flat-field, extraction, wavelength and flux calibration. Atmospheric extinction was corrected through mean coefficients derived for the LNA observatory. Table 2\footnote{Available electronically.} displays the line fluxes in a scale where $\mbox{F}(\mbox{H}{\beta}) = 100$, with reddening correction. A typical spectrum can be seen in figure \ref{pe115spec}, for the planetary nebula Pe 1-15.

\subsection{Interstellar extinction}

As pointed out by Escudero et al.~(\citeyear{escu04}), the application of the extinction curve of Fitzpatrick (\citeyear{fitz99}) has produced better results than the curve of Cardelli et al.~(\citeyear{ccm89}) for the interstellar extinction correction. Therefore we chose the former to correct the interstellar extinction, deriving E(B-V) from the observed Balmer ratio H$\alpha$/H$\beta$ and adopting the theoretical value $\mbox{H}\alpha/\mbox{H}\beta = 2.85$, with $\mbox{R}_{\mbox{V}} = 3.1$. This extinction curve is given by a seven-degree polynomial equation as follows:

\begin{eqnarray}
\left[\frac{\mbox{A}_{\lambda}}{\mbox{E(B-V)}}\right] & = & 0.00001+0.22707x + 1.95243x^2 - 2.67596x^3 + \nonumber \\ 
{} & & {} + 2.6507x^4 -1.26812x^5 + 0.27549x^6- \nonumber \\
{} & & {} - 0.02212x^7
\end{eqnarray}
with $x = 1/\lambda$ [$\mu\mbox{m}^{-1}$].


\section{Determination of chemical abundances}

\subsection{Physical parameters}

The physical parameters - electron densities and electron temperatures - were derived from optical emission lines. The electron density was determined from the sulfur line ratio [\ion{S}{II}] $\lambda$671.6/$\lambda$673.1 nm. For the electron temperature, we used both [\ion{O}{III}] $\lambda 436.3/ \lambda 500.7$ nm, and [\ion{N}{II}]  $\lambda 575.5/ \lambda 654.8$ nm line ratios, which gives two temperature zones: one for the low potential lines, and the other for high potential lines. 

Table \ref{tab_param} shows the physical parameters obtained for the observed PNe. Column 1 lists the PN G number, columns 2 and 3 the interstellar extinction E(B-V) with uncertainties (described in \S~\ref{errors}), columns 4-5 the electron density from [\ion{S}{II}] in $10^3 \mbox{cm}^{-3}$, and columns 6-7 and 8-9 the electron temperatures from [\ion{N}{II}] and [\ion{O}{III}], respectively, in units of $10^4\mbox{K}$ with uncertainties. Column 10 refers to the method used to obtain the electronic temperatures (see \S\ref{errors} for more details). 

For PN G004.2-0.59 the [O III] flux ratio resulted in a very high electron temperature, not typical for a planetary nebula. However, it is interesting to note that this ratio is very similar to that derived from the data of Exter et al.~(\citeyear{exter04}), what indicates that intrinsic properties of this nebula such as large density variations or the presence of shocked material could lead to an unusual flux ratio of the [OIII] lines, making them inappropriate to derive electron temperatures.  Additionally, this object does not have the [NII]5755 line, what makes impossible to obtain T([NII]). Spectra with better S/N as well as high quality, high resolution direct pictures of this object would be helpful to establish its nature. In view of this situation, we decided to keep the fluxes for this object but, since we cannot derive electron temperature, it is not included in the abundance analysis.

\addtocounter{table}{1}

\subsection{Ionic and elemental abundances}

Ionic abundances were calculated from the fits by Alexander \& Balick (\citeyear{alex97}), who provide convenient empirical relations for the determination of ionic abundances, obtained from numerical simulations. Ionic abundances for $\mbox{He}^+$ and $\mbox{He}^{++}$ were derived using the recombination coefficients from Pequignot et al.~(\citeyear{pequi91}). The $\mbox{He}^+$ abundance was also corrected of collisional effects using the correction terms from Kingdon \& Ferland (\citeyear{king95}). For the derivation of the $\mbox{O}^+$ abundance we chose the red pair of lines $\lambda 731.9+2.9$ nm, since they have better signal-to-noise in our spectra than the blue pair $\lambda 372.7+2.9$ nm counterpart, due to the greater efficiency of the instrumental set in the red region. Besides that, as discussed by Escudero et al.~(\citeyear{escu04}), there is a small difference between both determinations, with a tendency for smaller abundances when the blue lines are used. However, such difference is not larger than the errors involved in the determination of the abundances, so that we expect no measurable differences in the final oxygen abundance when using the red lines instead of the blue ones. 

For those objects where $\mbox{S}^{++}$ lines were not available we adopted the same technique used by Kingsburgh \& Barlow (\citeyear{kings94}) and Escudero et al.~(\citeyear{escu04}) to calculate the sulfur abundance. This technique consists of deriving the  $\mbox{S}^{++}$ abundance through a relation between the ratios $\mbox{S}^{++}$/$\mbox{S}^{+}$ and $\mbox{O}^{++}$/$\mbox{O}^{+}$. In this work we adopted the same relation used by Escudero et al.~(\citeyear{escu04}). The derived ionic abundances can be seen in table \ref{tab_ionic_abund}, where are also shown the errors for the ionic abundances obtained from a Monte Carlo simulation.

\addtocounter{table}{1}

Chemical abundances were calculated by means of ionization correction factors (ICFs), to account for unobserved ions of each element. The ICFs used were the same as those adopted by Escudero et al.~(\citeyear{escu04}), and were obtained from Kingsburgh \& Barlow (\citeyear{kings94}) for nitrogen, sulfur, and neon abundances; from Torres-Peimbert \& Peimbert (\citeyear{tpp77}) for the oxygen abundance; and from de Freitas Pacheco et al.~(\citeyear{fp93}) for argon. 

For helium in particular, we have the abundances of the ions $\mbox{He}^+$ and $\mbox{He}^{++}$. In agreement with the criterion defined by Torres-Peimbert \& Peimbert (\citeyear{tpp77}), there is no essential contribution from neutral helium in the total helium abundance when 
\begin{equation}
\label{neutral_helium}
\log \mbox{O}^+/\mbox{O} < -0.4, 
\end{equation}
and therefore  this component can be neglected. In this case, the helium total abundance can be written as
\begin{equation}
\frac{\mbox{He}}{\mbox{H}} = \frac{\mbox{He}^+}{\mbox{H}^+} + \frac{\mbox{He}^{++}}{\mbox{H}^+}.
\end{equation}

Table \ref{tab_abund} shows the chemical abundances and uncertainties obtained in this work in the notation $\epsilon(\mbox{X}) = \log(\mbox{X/H})+12$, where X denotes N, S, O, Ar, and Ne. For He, the He/H is given instead. When the condition expressed by equation \ref{neutral_helium} is not satisfied, a $\star$ symbol is displayed in front of the PN G number. In these cases, the helium abundances are lower limits for the total helium abundances and must be considered carefully. In some cases, the method used to calculate the errors and mean abundances (see section \ref{errors}) did not converge, and the abundances and errors were replaced by the mean and the standard deviation obtained from the independent measures for each object. These abundances are indicated with an * in table \ref{tab_abund}.

\addtocounter{table}{1}

\subsection{Errors \label{errors}}

\begin{figure}[!ht]
   \centering
   \includegraphics[width=9cm]{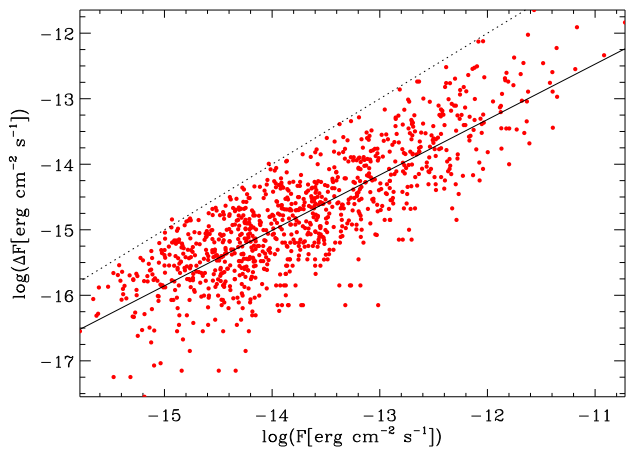}
   \caption{Errors in line fluxes as a function of the fluxes. Dotted line represents $\Delta$F = F, and the continuous line is the linear fit to the data.}
    \label{flux_errors}%
\end{figure}

In order to determine errors in the physical parameters and abundances, gaussian noise was added to the observed spectra by means of a Monte Carlo simulation, where  each line flux was varied randomly 500 times within its respective error interval. These error intervals were estimated from a relation between the errors in fluxes and fluxes that were obtained from a linear fit in the data as shown in Figure \ref{flux_errors}. In this figure, the horizontal axis shows the mean reddened fluxes for each line of each object and the vertical axis shows the errors in line fluxes, which are the standard deviations calculated from the independent line flux measurements for each object. 

The linear fit gives:

\begin{equation}
 \log (\Delta\mbox{F}) =( -3.2 \pm 0.2) + (0.85 \pm 0.02)\log (\mbox{F}). 
\end{equation}

The colour excesses were also varied randomly within their error interval, which were estimated from the standard deviation calculated for each measure. Final abundances and electron temperatures were adopted as the peak of a gaussian fit to the 
histogram of the 500 random generated values and the errors were adopted as half of the FWHM of the gaussian profile fitted to each histogram, except for densities, whose value distribution is not gaussian. Densities and errors were estimated from the mean and standard deviation calculated for the different measurements of each object, respectively.

Average errors for the whole sample of abundances are shown in table \ref{tab_errors}, where the columns indicate the chemical element and the error associated with it. These errors were obtained from the mean of the standard deviations of the chemical abundances for each element. We have to stress the fact that these errors take into account only the influence of line flux uncertainties in the chemical abundances. The major source of errors in our determination of chemical abundances is due to uncertainties in the ICFs, hence the errors derived here only measure the dispersion of the observational data.
    
\begin{table}[ht!]
\caption{Mean errors for the abundances in dex. For helium, the uncertainty in the He/H ratio is given.} 
\label{tab_errors}      
\centering   
\small                  
\begin{tabular}{c  c  c  c  c  c  c }        
\hline \hline
Element  &  He          &     N              &        S            &       O            &        Ar           &       Ne           \\
\hline
Error    & $\pm 0.021$  & $\pm 0.14$         &    $\pm 0.15$       &    $\pm 0.14$      &     $\pm 0.16$      &     $\pm 0.13$     \\                      
\hline \hline
\end{tabular}
\end{table}


\section{Results}
\label{discussion_data}

In order to check the consistency of the data presented in this work, as given in table \ref{tab_abund}, we analysed some statistical properties of the sample and compared it with chemical abundances taken from the literature. We searched for chemical abundances of PNe located in the bulge and inner-disk of the Galaxy in the following works: Ratag et al.~(\citeyear{ratag97}), hereafter RPDM97; Exter et al.~(\citeyear{exter04}), hereafter EBW04; G\'orny et al.~(\citeyear{gorny04}), hereafter GSEC04; and Escudero et al.~(\citeyear{escu04}), hereafter ECM04. All these works have the same region of interest, and, besides that, they have significant and homogeneous samples.

RPDM97 derived abundances for a sample of 45 bulge PNe based on theoretical photoionization models used to account for individual ICF for each PN. They also reanalysed the data for 50 bulge PNe previously published.

EBW04 published chemical abundances for 45 bulge PNe using the empirical method, as in this work. They use ICFs from Kingsburgh \& Barlow (\citeyear{kings94}).

ECM04 observed 57 bulge PNe using the empirical method to derive the abundances. They adopted the blue line pair to derive the $\mbox{O}^+$ abundance. The ICFs used by them are as in the present work. 

GSEC04 observed 44 PNe towards the bulge and the abundances were derived using the empirical method. They used the [\ion{N}{II}] temperature for the ions of low ionization level and the [\ion{O}{III}] temperature for those with high ionization level, to derive the abundances. They adopted as $\mbox{O}^+$ abundance the mean between the abundances obtained from the [\ion{O}{II}] $\lambda 372.7$ and [\ion{O}{II}] $\lambda 732.0, 733.0$ nm lines.

\subsection{Abundance distributions}

\begin{figure}[!ht]
	\centering
	\includegraphics[width=9cm]{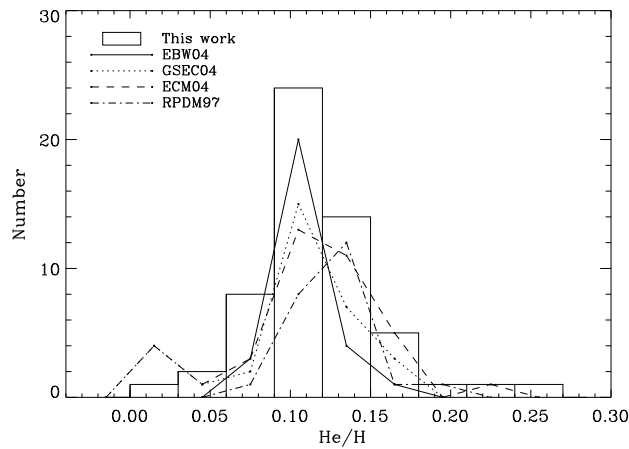}
	\caption{Comparison between distribution of helium abundances derived in this work (boxed histogram) and the abundances taken from the literature (lines).}	\label{helium}
\end{figure}

\begin{figure}[!ht]
	\centering
	\includegraphics[width=9cm]{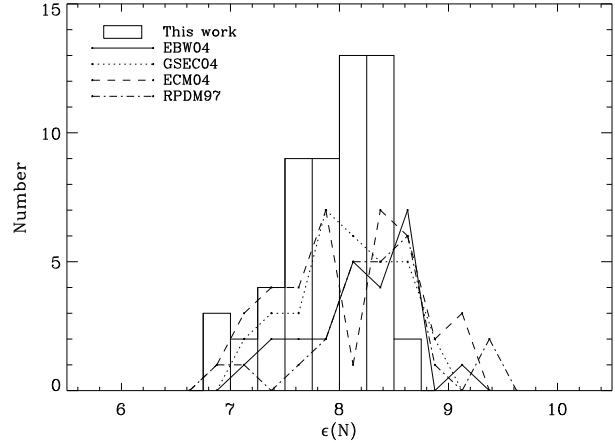}
	\caption{The same as figure \ref{helium} for nitrogen.}
	\label{nitrogen}
\end{figure}

\begin{figure}[!ht]
	\centering
	\includegraphics[width=9cm]{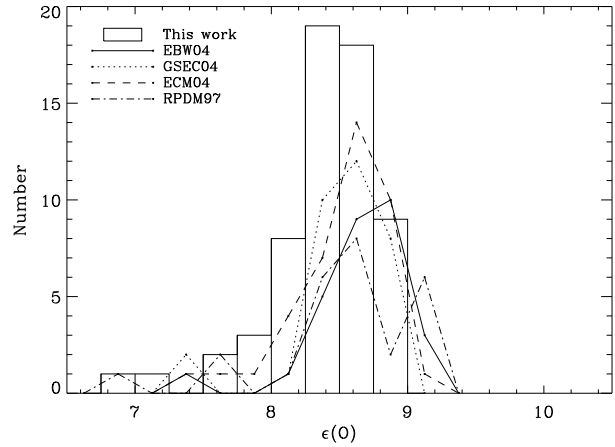}
	\caption{The same as figure \ref{helium} for oxygen.}
	\label{oxygen}
\end{figure}

\begin{figure}[!ht]
	\centering
	\includegraphics[width=9cm]{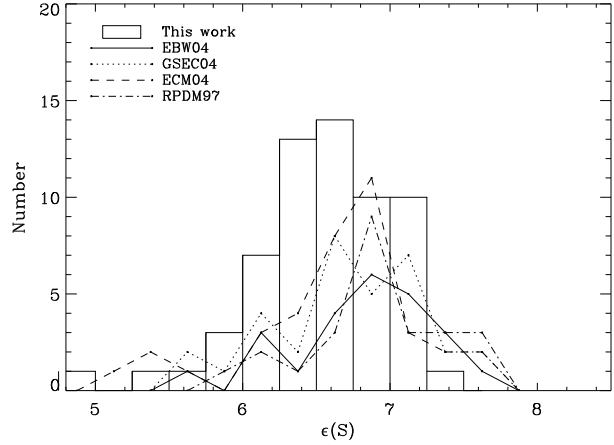}
	\caption{The same as figure \ref{helium} for sulfur.}
	\label{sulfur}
\end{figure}

\begin{figure}[!ht]
	\centering
	\includegraphics[width=9cm]{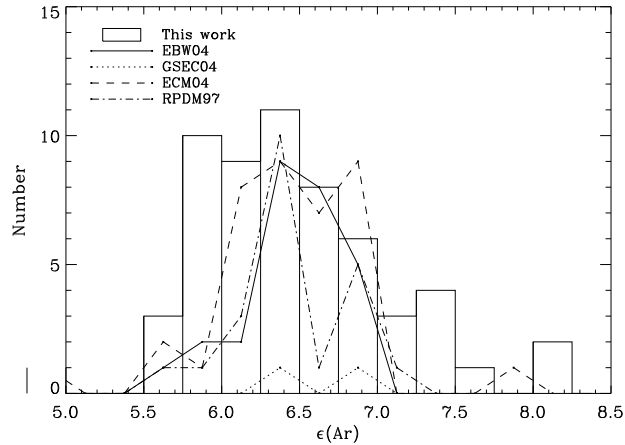}
	\caption{The same as figure \ref{helium} for argon.}
	\label{argon}
\end{figure}

\begin{figure}[!ht]
	\centering
	\includegraphics[width=9cm]{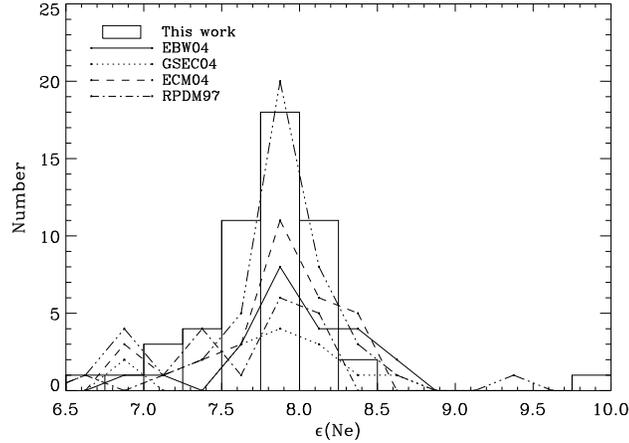}
	\caption{The same as figure \ref{helium} for neon.}
	\label{neon}
\end{figure}

Figures \ref{helium} to \ref{neon} show the distribution of the chemical abundances obtained in this work (boxed histogram) compared with data from the literature (lines). Each line is from a different work as indicated at the top left in each figure. 

Since helium and nitrogen abundances are modified by the evolution of intermediate mass stars (IMS), the histograms in figures \ref{helium} and \ref{nitrogen} show the results of this evolution coupled to the chemical evolution of the Galaxy. For helium in particular, the histogram shows a wide distribution, which results from the mass and age ranges of the progenitor stars that originate the PNe. This behaviour is seen both in our data as well as in data from the literature. It is important to note the good agreement between the data from different authors and the present work as shown by these histograms. 

In agreement with theories of stellar evolution and nucleosynthesis (Lattanzio \& Forestini \citeyear{lat99}), we do not expect significant changes in the abundances of the $\alpha$-elements (O, S, Ar, Ne), so that these abundances reflect the abundances of the interstellar medium at the progenitor formation epoch, indicating the chemical evolution of the Galaxy. 

The histograms for oxygen, sulfur, argon, and neon, are displayed in figures \ref{oxygen}, \ref{sulfur}, \ref{argon}, and \ref{neon}, respectively. In figure \ref{oxygen}, oxygen shows a systematic lower abundance than the data from the literature by approximately 0.2 dex, which is larger than the expected errors for the oxygen abundance obtained in this work. However, the bulge is formed by different populations, mostly by stars with ages $10 \pm 2.5$ Gyr (Zoccali et al.~\citeyear{zoc03}). On the other hand, there are evidences for an younger population formed by OH/IR stars, AGB variables, etc, that appear to be set in the galactic plane (van Loon et al.~\citeyear{vanloon03}). Such age distribution results in a wide abundance distribution for oxygen and $\alpha$-elements in general. It can be seen that there is a significant number of PNe in the range 8 to 9 dex, showing that the progenitor stars of the bulge PNe were formed in different epochs, suggesting a scenario where the bulge was formed in a diversity of epochs, as discussed by Costa \& Maciel (\citeyear{costa06}).

For sulfur, argon and neon (figures \ref{sulfur}, \ref{argon} and \ref{neon}, respectively) the distributions of the abundances are very similar to the data from the literature, in the sense that they are very wide, and the peaks of the distributions match each other. It is worth to note that the distributions of argon are bimodal for RPDM97 and ECM04 data.

\subsection{Comparison of individual abundances}

In this work, 21 out of 54 objects have their abundances already published in the literature. The remaining 33 objects have their abundances published for the first time. Table \ref{tab_comp} shows a comparison between our data (left column for each element) and data from the literature (right column for each element). The references are shown in the last column, where the abbreviations denote CKA96 (Cuisinier et al.~\citeyear{cuisin96}), GKA07 (Girard et al.~\citeyear{gir07}), P91 (Perinotto~\citeyear{peri91}), CGSB09 (Chiappini et al.~\citeyear{chiap09}), PMS04 (Perinotto et al.~\citeyear{peri04}), WL07 (Wang \& Liu \citeyear{wang07}). P91 compiled a catalogue of chemical abundances for 209 PNe, taken from the literature. CAK96 gives chemical abundances for 62 PNe derived by the same empirical method as used in this work but the ICFs are from their own model, except for N, S and Ne, whose ICFs are from Kingsburgh \& Barlow (\citeyear{kings94}), as in this work. GKA07 analyse chemical properties for 48 PNe around central stars of spectral types [WC], [WO]. The ICFs are from Aller (\citeyear{aller84}) which, except for argon, are the same we used. Recently CGSB09 published chemical abundances for 245 objects belonging to the bulge and inner disk of the Galaxy, from which 90 PNe were observed with 4m-class telescopes by G\'orny et al.~(\citeyear{gorny09}), and the remain have their chemical abundances recalculated from an empirical method described in  G\'orny et al.~(\citeyear{gorny09}). They have the advantage over our observations the size of the telescope, but have as disadvantage the fact that some spectra were derredened by using the ratio H$\alpha$/H$\gamma$ instead of the usual ratio H$\alpha$/H$\beta$. This can introduce errors in the derredened fluxes, which are propagated to the chemical abundances. PMS04 did a reanalysis of all chemical abundances published so far in a very homogeneous way. The ICFs used are as in the present work, but the extinction curve used is from Mathis (\citeyear{mathis90}), while here we made use of those from Fitzpatrick (\citeyear{fitz99}). Finally, WL07 give chemical abundances for 25 galactic bulge PNe and 6 from the galactic disk determined from both collisional excitation lines  and optical recombination lines, by solving level populations of the emitting ions. The ICFs are the same as adopted in Wesson et al.~(\citeyear{wesson05}), which for S and Ne are the same as used in this work.  

\begin{table*}[!ht]
\setlength{\tabcolsep}{1pt}
\centering
\scriptsize
\caption{Comparison between abundances from this work (left column for each element) and abundances from the literature (right column) for PNe with abundances previously published.}    
                    
\begin{tabular}{l  c  c  c  c  c  c  c  c  c  c  c  c  c}
\hline \hline
  & \multicolumn{2}{c}{He/H} & \multicolumn{2}{c}{$\epsilon$(N)} &  \multicolumn{2}{c}{$\epsilon$(S)} &  \multicolumn{2}{c}{$\epsilon$(O)}  & \multicolumn{2}{c}{$\epsilon$(Ar)} & \multicolumn{2}{c}{$\epsilon$(Ne)} & \\[-.25cm]    
  & \multicolumn{2}{c}{\rule{2cm}{.01cm}}& \multicolumn{2}{c}{\rule{2cm}{.01cm}}&\multicolumn{2}{c}{\rule{2cm}{.01cm}}&\multicolumn{2}{c}{\rule{2cm}{.01cm}}& \multicolumn{2}{c}{\rule{2cm}{.01cm}}& \multicolumn{2}{c}{\rule{2cm}{.01cm}}& \\[-.1cm]
Name & \scriptsize{our} & \scriptsize{lit.} & \scriptsize{our} & \scriptsize{lit.} & \scriptsize{our} & \scriptsize{lit.} & \scriptsize{our} & \scriptsize{lit.} & \scriptsize{our} & \scriptsize{lit.} & \scriptsize{our} & \scriptsize{lit.} & Ref.\\
\hline\\[-.25cm]

   H1-8			   & 0.073		    & 0.152 & 8.19		    & 8.68 & 6.55                  & 7.04 & 8.86		  & 8.76 & 7.28                  & 6.80 & 8.29                  & --   & CGSB09 \\[.1cm]

   H 1-9		   & 0.063		    & 0.073 & 6.82                  & 7.41 & 6.11       	   & 6.51 & 7.73                  & 8.30 & 6.19                  & 5.85 & 6.71                  & 6.84 & CGSB09 \\[.1cm]

   \multirow{4}{*}{H 1-17} & \multirow{4}{*}{0.112} & 0.100  & \multirow{4}{*}{8.07}& 8.32 & \multirow{4}{*}{6.81} & 6.78 & \multirow{4}{*}{8.36} & 8.61 & \multirow{4}{*}{6.34} & 6.42 & \multirow{4}{*}{7.55} & --   & RPDM97\\
                           &                        & 0.083  &                      & 8.08 &                       & 6.62 &                       & 8.32 &                       & 6.09 &                       & --   & CMKAS00\\
                           &                        & 0.086  &                      & 7.73 &                       & 6.52 &                       & 8.35 &                       & 6.34 &                       & 7.90 & EBW04 \\
                           &	                    & 0.118 &                       & 8.04 &                       & 6.82 &                       & 8.58 &                       & 6.38 &                       & 8.00 & CGSB09\\[.1cm]                     
   \multirow{2}{*}{H 1-60} & \multirow{2}{*}{0.114} & 0.098 &  \multirow{2}{*}{8.02}& --   &  \multirow{2}{*}{6.66}& 6.93 & \multirow{2}{*}{8.45} & 8.56 &  \multirow{2}{*}{6.00}& 6.41 &  \multirow{2}{*}{7.80}& 7.90 & RPDM97 \\
                           &	                    & 0.117 &                       & 7.46 &                       & 6.50 &                       & 8.58 &                       & 6.13 &                       & --   & CGSB09 \\[.1cm]  
   
   \multirow{2}{*}{H 2-1}  & \multirow{2}{*}{0.059} & 0.038 & \multirow{2}{*}{7.00} & 6.73 & \multirow{2}{*}{5.79} & 6.06 & \multirow{2}{*}{7.92} & 7.71 & \multirow{2}{*}{6.52} & 5.35 & \multirow{2}{*}{7.08} & --   & PMS04 \\       
                           &                        & 0.041 &                       & 7.16 &                       & 5.97 &                       & 7.76 &                       & 5.42 &                       & --   & CGSB09\\[.1cm]
   
   \multirow{2}{*}{H 2-10} & \multirow{2}{*}{0.090} & 0.089 & \multirow{2}{*}{7.72} & 7.87 & \multirow{2}{*}{6.42} & 6.74 & \multirow{2}{*}{8.24} & 8.62 & \multirow{2}{*}{5.76} & 5.78 & \multirow{2}{*}{7.79} & --   & CMKAS00 \\      
                           &	                    & 0.103 &                       & 7.80 &                       & 6.62 &                       & 8.54 &                       & 5.84 &                       & --   & CGSB09  \\[.1cm]

   H 2-45		   & 0.096       	    & 0.103 & 7.40		    & 7.85 & 6.28		   & 6.53 & 8.27                  & 8.39 & 5.62           	 & 5.84	& 7.57                  & --   & CGSB09 \\[.1cm]

   Hf 2-1	           & 0.127                  & 0.135 & 8.66		    & 8.71 & 7.02                  & 7.19 & 8.83                  & 8.73 & 6.88                  & 6.53 & 8.10                  & --   & CGSB09 \\[.1cm] 

   \multirow{2}{*}{IC4699} & \multirow{2}{*}{0.079} & 0.098 & \multirow{2}{*}{--}   & 7.34 & \multirow{2}{*}{6.15} & 6.34 & \multirow{2}{*}{8.50} & 8.49 & \multirow{2}{*}{5.62} & 6.28 & \multirow{2}{*}{7.78} & 7.79 & WL07  \\    
                           &                        & 0.101 &                       & 7.69 &                       & 6.49 &                       & 8.50 &                       & 6.37 &                       & 7.81 & CGSB09\\[.1cm] 
   M 1-39	           & 0.089                  & 0.067 & 8.21                  & 8.22 & 7.03                  & 7.02 & 8.70                  & 8.61 & 7.38                  & 6.45 & --                    & --   & CGSB09\\[.1cm]

   \multirow{2}{*}{M 1-45} & \multirow{2}{*}{0.017} & --    & \multirow{2}{*}{8.23} & 8.41 & \multirow{2}{*}{6.85} & 6.82 & \multirow{2}{*}{8.49} & 8.68 & \multirow{2}{*}{6.89} & -- & \multirow{2}{*}{--} & -- & CAK96  \\
                           &                        & 0.112 &                       & 8.30 &                       & 6.96 &                       & 8.73 &                       & -- &                        & -- & RPDM97\\[.1cm]

   M 1-46                  &  0.104                 & 0.080 & 7.75                  & 7.95 & 6.26                  & 7.13 & 8.97                  & 8.87 & 8.15                  & 6.53 & 7.62                  & --    & GKA07 \\[.1cm]
   M 1-47                  &  0.096                 & 0.112 & --                    & 8.15 & 6.34                  & --   & 8.47                  & 8.51 & 5.73                  & 5.85 & 7.76                 & -- & CKA96 \\[.1cm] 
   \multirow{2}{*}{M 1-60} & \multirow{2}{*}{0.127} & 0.117 & \multirow{2}{*}{8.50} & 8.73 & \multirow{2}{*}{6.86} & 7.4  & \multirow{2}{*}{8.52} & 9.06 & \multirow{2}{*}{6.38} & 6.74 & \multirow{2}{*}{8.00} & --    & CAK96\\    
                           &                        & 0.117 &                       & 8.96 &                       & 7.21 &                       & 8.78 &                       & 6.67 &                       &  8.22 & GKA07\\[.1cm]

   \multirow{2}{*}{M 2-21} & \multirow{2}{*}{0.110} & 0.120 & \multirow{2}{*}{7.55} & 7.95 & \multirow{2}{*}{6.78} & --   & \multirow{2}{*}{8.47} & 8.49 & \multirow{2}{*}{5.92} & --   & \multirow{2}{*}{7.38} & 7.77  & P91    \\    
                           & 	                    & 0.119 &                       & 7.75 &                       & 6.30 &                       & 8.41 &                       & 5.83 &                       & 7.69  & CGSB09 \\[.1cm]                     
   \multirow{3}{*}{M 2-22} & \multirow{3}{*}{0.147} & 0.140 & \multirow{3}{*}{8.95} & 8.48 & \multirow{3}{*}{6.75} & --   &  \multirow{3}{*}{8.27}& 8.53 & \multirow{3}{*}{6.37} & --   & \multirow{3}{*}{7.84} & --    & P91    \\
                           &                        & 0.148 &                       & 8.45 &                       & 7.03 &                       & 8.59 &                       & 6.66 &                       & 8.30  & RPDM97 \\
                           &	                    & 0.166 &                       & 8.67 &                       & 6.97 &                       & 8.63 &                       & 6.56 &                       & --    & CGSB09 \\[.1cm]                     
   \multirow{2}{*}{M 3-23} & \multirow{2}{*}{0.118} & 0.122 & \multirow{2}{*}{8.27} & --   & \multirow{2}{*}{6.99} & --   &  \multirow{2}{*}{8.79}& 8.64 & \multirow{2}{*}{6.76} & --   & \multirow{2}{*}{8.14} & 7.92  & EBW04 \\         
                           &	                    & 0.125 &                       & 8.50 &                       & 7.11 &                       & 8.63 &                       & 6.54 &                       & 7.95  & CGSB09\\[.1cm]                     

	M 3-24		   & 0.149		    & 0.165 & 8.22   		    & 8.34 & 6.74		   & 6.73 &  8.48		  & 8.51 &  6.45		 & 6.41	&  7.96			& 8.01	& CGSB09\\[.1cm]

   \multirow{2}{*}{M 3-29} & \multirow{2}{*}{0.088} & 0.100 & \multirow{2}{*}{7.56} & 7.98 & \multirow{2}{*}{6.22} & 6.70 & \multirow{2}{*}{8.45} & 8.51 & \multirow{2}{*}{6.24} & 5.89 & \multirow{2}{*}{8.08} & 7.85 & WL07  \\                            
                           &                        & 0.101 &                       & 8.07 &                       & 6.70 &                       & 8.51 &                       & 6.00 &                       & 7.88 & CGSB09\\[.1cm]

   Pe 2-13	           & 0.145                  & 0.156 & 7.80                  &  --  & 7.00		   & --	  & 8.40                  & 8.67 & 6.59                  & 6.55 & 7.58                  & --   & CGSB09 \\[.1cm]

   Vd 1-8	           & 0.099                  & 0.115 & 7.70                  & 7.90 & 6.23                  & 6.56 & 8.13                  & 8.41 & 5.80                  & 6.21 & 7.44                  & --   & CGSB09\\

\hline 
\label{tab_comp}                                  
\end{tabular}
\end{table*}

\begin{figure}[!ht]
  \centering
 \includegraphics[width=9cm]{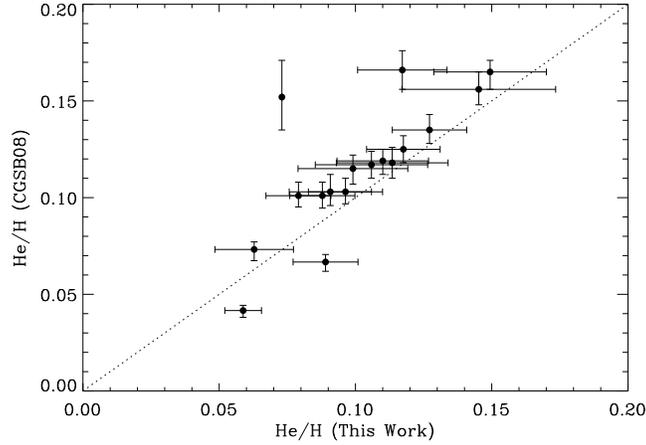}
 \caption{Comparison between abundances from this work and CGSB09 for helium with error bars. The dashed line represents the equality between the data.\label{he_our_cgsb09}}
\end{figure}

\begin{figure}[!ht]
 \centering
 \includegraphics[width=9cm]{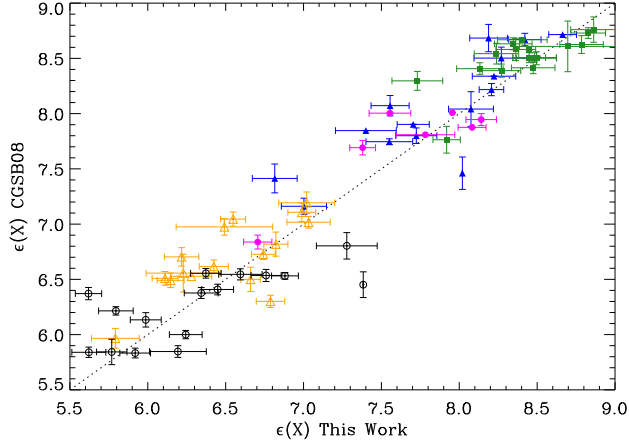}
 \caption{The same as figure \ref{he_our_cgsb09} but for other elements. The abundances are in units of $\log(\mbox{X/H})$ where X stands for nitrogen (filled triangles), oxygen (filled squares), neon (filled circles), sulfur (open triangles), argon (open circles).\label{elem_our_cgsb09}}
\end{figure}

In order to verify the dispersion of the results, we computed the difference between the abundances obtained in this work and those from the literature. The helium abundances from this work differ of the abundances from literature by 0.01, which is lower than our error estimate for the helium abundances of this work. For the other elements, the differences are 0.16, 0.21, 0.14, 0.24, and 0.21 dex, for N, S, O, Ar, and Ne, respectively. These differences are similar to the errors estimated for these abundances, and are the result of the differences between the methods employed to obtain the chemical abundances, such as ICFs, as well as the errors associated to different observation and reduction processes. In particular, we can compare our data to those from CGSB09 directly, since there are enough objects in common between the two samples. Figures \ref{he_our_cgsb09} and \ref{elem_our_cgsb09} show a comparison between abundances from this work and CGSB09 for helium. Abundances from CGBS09 display a systematic tendency to higher values compared to our data. Nonetheless the difference is not superior to the expected errors for the abundances.

\subsection{Abundance correlations \label{abund_corr}}
	
Correlations between the chemical abundances for the different elements are an important tool to understand the evolution of the central stars of PN (CSPN). In particular, the correlations between elements not produced by the progenitor stars give important information about the nucleosynthesis of massive stars and the formation and evolution of the Galaxy, as described by Ballero et al.~(2007).

\begin{figure}[!ht]
	\centering
	\includegraphics[width=9cm]{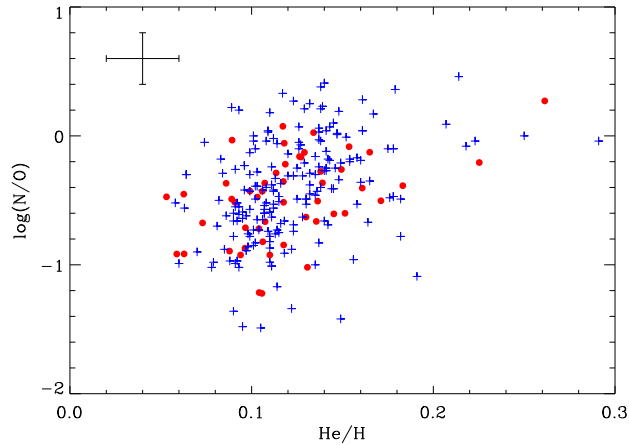}
	\caption{$\log(\mbox{N}/\mbox{O})$ as a function of He/H. The crosses are data from the literature and filled circles are our data. The error bar at the upper left corner indicates the mean errors for the abundances.}
	\label{logno_he}
\end{figure}

In figure \ref{logno_he} we show the correlation between $\log(\mbox{N}/\mbox{O})$ and He/H for our data and from the works previously mentioned. As discussed earlier, helium and nitrogen abundances are modified during the evolution of the progenitor star, and PNe with high abundances of helium and nitrogen are originated from massive stars, so that the correlation between these elements must be positive. From figure \ref{logno_he} it can be seen that the correlation between $\log(\mbox{N}/\mbox{O})$ and helium is positive, although there is no tight correlation between these quantities. Indeed, excluding the helium abundances lower than 0.050, which are not realistic, and probably indicate the presence of neutral helium in these nebulae, the linear Pearson correlation coefficient of our data is 0.47, showing a small correlation. The whole sample, which consists of literature and our data, shows a correlation coefficient 23\% lower compared with our data. It is important to note that both ours and literature data show a large spread in this correlation, which is probably related to distinct efficiencies in the mixing episodes occurring along the evolution for stars with different masses. It is expected that nitrogen enhancement would not be so high in non-type I PNe, which represents 80 \% of the PNe population in the Galaxy (Peimbert \& Serrano \citeyear{peimb80}). In figure \ref{logno_he} it is possible to see that most objects show a low N/O ratio, except for a small number of PNe with $\log(\mbox{N}/\mbox{O})$ ratio close to 0.5. These PNe could be originated from massive stars, pointing to a recent star formation. Cuisinier et al.~(\citeyear{cuisin00}) showed that bulge PNe have lower N/O ratio compared with disk PNe. Therefore, those PNe with high N/O ratio could be located in the transition between the disk and bulge. 

\begin{figure}[!ht]
	\centering
	\includegraphics[width=9cm]{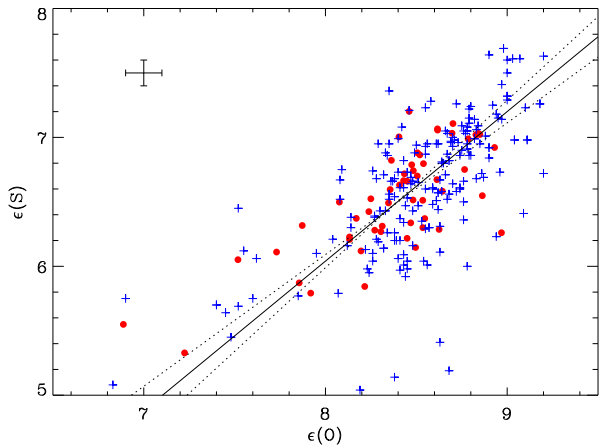}
	\caption{Correlation between sulfur and oxygen abundances from our results (filled circles) and the literature (crosses). Mean errors are shown at the upper left corner. The continuous line is a linear bisector fit to the data, while dashed lines are the one-sigma confidence level obtained from the uncertainties in the fit parameters.}
	\label{sulf_oxy}
\end{figure}

\begin{figure}[ht!]
	\centering
	\includegraphics[width=9cm]{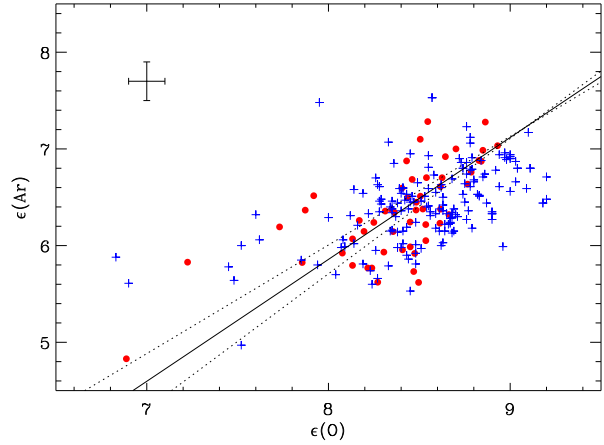}
	\caption{The same as figure \ref{sulf_oxy} for argon and oxygen.}
	\label{arg_oxy}
\end{figure}

\begin{figure}[!ht]
	\centering
	\includegraphics[width=9cm]{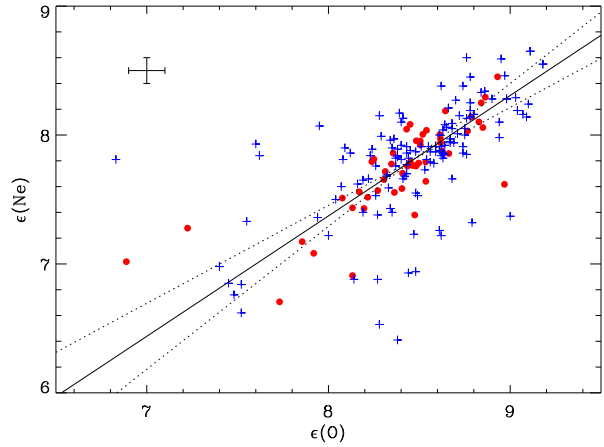}
	\caption{The same as figure \ref{sulf_oxy} for neon and oxygen.}
        \label{neon_oxy}  
\end{figure}

Figures \ref{sulf_oxy}, \ref{arg_oxy}, and \ref{neon_oxy} show the correlations between S, Ar and Ne with O. The symbols are as in figure \ref{logno_he}. Since these elements are produced by the same process, and their abundances do not change significantly in IMS, a positive correlation is expected between the sulfur, argon, neon and oxygen abundances in PNe. In these figures positive correlations are observed, with linear correlation coefficients of 0.60, 0.68, and 0.78 for sulfur, argon, and neon, respectively. These correlations indicate a medium to large correlation. Concerning our data, excluding the helium abundances lower than 0.050 since they are not realistic, as discussed before, the slope and the y-intercept of a bisector method for the correlation between sulfur, argon, and neon with oxygen are, in this order, $1.2\pm0.2$ and $(-0.3\pm0.1)\times10$, $1.3\pm0.2$ and $(-0.4\pm0.1)\times10$, $0.9\pm0.1$ and $(-0.0\pm0.1)\times10$. These slopes differ from the whole sample by -4\%, 20\%, and -15\% for sulfur, argon, and neon, respectively. From these results we can see that a linear correlation between the $\alpha$-elements with oxygen with a slope close to unity is a good approximation. The linear correlations between $\alpha$-elements and oxygen seen throughout this section suggest that these elements are in lockstep in PNe, so that modifications (if any) during the evolution of the progenitor star are small. Again, there is a generally good agreement between our new abundances and those from the literature.

\section{Summary and Conclusion}

In summary, this work reports an important result concerning PNe and the chemical evolution of the Galaxy. 

We present the analysis of chemical abundances of a PNe sample located towards the galactic bulge. New chemical abundances were derived through spectrophotometric observations made at the 1.60 m telescope of the LNA-Brazil, comprising the elements He, N, S, O, Ar, and Ne. 54 PNe were considered, among which 33 objects have their abundances derived for the first time. A comparison between the chemical abundances from this work and abundances obtained from the literature was performed. The analysis shows that the distributions of abundances are similar but not identical. Some objects of this work are listed in other investigations and a direct comparison between these abundances shows that the differences are of the order of 0.2 dex, indicating that the distinct methods used to derive the abundances are the main source of this difference. With the present results, we intend to enlarge the number of planetary nebulae with accurate chemical abundances, providing a large and homogeneous set of chemical abundances, contributing to the understanding of this stage of star evolution as well as the study of the chemical evolution of the inner Galaxy. 


\bigskip
{\it Acknowledgements.
      Part of this work was supported by the Brazilian agencies \emph{FAPESP} and \emph{CNPq}. O.C. would like to acknowledge FAPESP
for his graduate fellowship (processes 05/03194-4 and 07/07704-2).
}

\newpage

\setcounter{table}{0}

\begin{longtable}{llcccl}
\caption{\label{tab_log} Log of the observations.}\\
\hline\hline
PN G & Name & RA (J2000) & DEC (J2000) & Date of Obs. & Exp. Time (s) \\
\hline
\endfirsthead
\caption{continued.}\\
\hline\hline
PN G & Name & RA (J2000) & DEC (J2000) & Date of Obs. & Exp. Time (s) \\
\hline
\endhead
\hline
\endfoot

000.7-02.7		&		M 2-21			&		17 58 09.57		&		-29 44 20.10		&	Jun 24, 07	&		$2\times1200$	\\	
000.9-04.8		&		 M 3-23   		&		18 07 06.15		&		-30 34 17.00		&	Jun 21, 07	&		$3\times1200$	\\	
004.0-11.1 		&	 	M 3-29  		&	 	18 39 25.77 		&	 	-30 40 36.70		&	Jun 24, 06	&	 	$2\times1200$	\\	
004.2-04.3      	&	       H 1-60 			&	 	18 12 25.16		&		-27 29 13.00		&	Jun 21, 07	&		$2\times1200$	\\	
004.2-05.9		&		  M 2-37 		&		18 18 38.35		&		-28 08 01.00		&	Jun 24, 07	&		$1\times1200$	\\	
005.2-18.6 		&	 	StWr 2-21 		&	 	19 14 23.33 		&	 	-32 34 16.70		&	Jun 21, 06	&	 	$3\times900$	\\	
005.5-2.5		&		M 3-24			&		18 07 53.91		&		-25 24 02.71		&	Jun 24, 07 	&		$3\times900$    \\ 
006.4-04.6 		&		  Pe 2-13		&		18 18 13.36		&		-25 38 08.90		&	Jun 22/23, 07 	&		$2\times1800$	\\	
006.8-03.4 		&		  H 2-45 		&		18 14 28.84		&		-24 43 38.30		&	Jun 23, 07	&		$2\times900$	\\	
007.0+06.3 		&		  M 1-24 		&		17 38 11.59		&		-19 37 37.60		&	Jun 24, 07	&		$2\times1200$	\\	
010.7+07.4		&	 	Sa 2-230 		&	 	17 42 02.01 		&	 	-15 56 07.50		&	Jun 24, 06	&	 	$2\times1800$	\\	
011.0-05.1		&		M 1-47			&		18 29 11.15		&		-21 46 53.40 		&	Jun 21, 07	&	 	$2\times900$	\\	
011.3+02.8 		&	 	Th 4-11 		&	 	18 00 08.82 		&	 	-17 40 43.30		&	Jun 24, 06	&	 	$3\times600$	\\	
011.7-06.6 		&	 	M 1-55  		&	 	18 36 42.55 		&	 	-21 48 59.10		&	Jun 23, 06	&	 	$2\times1200$	\\	
012.6-02.6		&		M 1-45			&		18 23 07.98		&		-19 17 05.30		&	Jun 21, 07	&		$2\times1200$	\\	
013.8-07.9 		&	 	Pc 21   		&	 	18 45 35.22 		&	 	-20 34 58.30		&	Jun 21, 06	&	 	$2\times1200$	\\	
015.9+03.3 		&	 	M 1-39  		&	 	18 07 30.70 		&	 	-13 28 47.60		&	Jun 23, 06	&	 	$3\times600$	\\	
016.4-01.9		&		M 1-46			&		18 27 56.34		&		-15 32 54.40		&	Jun 23, 07	&		$3\times600$	\\	
019.7-04.5		&		M 1-60			&		18 43 38.11		&		-13 44 48.60		&	Jun 23, 07	&		$3\times600$	\\	
021.8-00.4     		&		 M 3-28   		&		18 32 41.29		&		-10 05 50.00		&	Jun 22, 07	&		$2\times1200$	\\	
023.0+04.3 		&	 	MA 3    		&	 	18 17 49.38 		&	 	-06 48 21.50		&	Jun 24, 06	&	 	$3\times1200$	\\	
023.3-07.6 		&	 	MaC 1-16 		&	 	19 01 21.77 		&	 	-11 58 20.00		&	Jun 21, 06	&	 	$3\times1200$	\\	
023.8-01.7     		&		 K 3-11       		&		18 41 07.31		&		-08 55 59.00		&	Jun 22, 07	&		$2\times1800$	\\	
024.1+03.8 		&	 	M 2-40  		&	 	18 21 23.85 		&	 	-06 01 55.80		&	Jun 24, 06	&	 	$2\times1200$	\\	
025.9-02.1  		&		 Pe 1-15 		&		18 46 24.48		&		-07 14 34.60		&	Jun 22, 07	&		$2\times1200$	\\	
335.4-01.1		&		He 2-169		&		16 34 13.33		&		-49 21 13.20		&	Jun 23, 07	&		$2\times1200$	\\	
335.9-03.6 		&	 	Mewe 1-7 		&	 	16 47 57.07 		&	 	-50 42 48.30		&	Jun 21, 06	&	 	$3\times1200$	\\	
336.2+01.9 		&	 	Pe 1-6  		&	 	16 23 54.31 		&	 	-46 42 15.30		&	Jun 22, 07	&	 	$2\times1200$	\\	
336.3-05.6     		&		He 2-186		&		16 59 36.06		&		-51 42 06.50		&	Jun 25, 07	&		$3\times600$	\\	
336.9+08.3 		&	 	StWr 4-10 		&	 	16 02 13.04 		&	 	-41 33 35.90		&	Jun 23, 06	&	 	$2\times1200$	\\	
338.8+05.6		&		He 2-155		&		16 19 23.10		&		-42 15 36.00		&	Jun 21, 07	&		$3\times900$	\\	
340.9-04.6 		&	 	Sa 1-5  		&	 	17 11 27.37 		&	 	-47 25 01.60		&	Jun 23, 06	&	 	$2\times900$	\\	
342.9-04.9 		&	 	He 2-207		&	 	17 19 32.97 		&	 	-45 53 16.70		&	Jun 23, 06	&	 	$3\times900$	\\	
343.0-01.7 		&	 	Vd 1-9  		&	 	17 05 38.30 		&	 	-43 56 18.00		&	Jun 21, 06	&	 	$2\times1200$	\\	
344.2-01.2		&		 H 1-6       		&		17 06 58.87		&		-42 41 09.75		&	Jun 23, 07	&		$2\times1800$	\\	
344.4+02.8 		&	 	Vd 1-5  		&	 	16 51 33.57 		&	 	-40 02 56.00		&	Jun 21, 06	&	 	$2\times1200$	\\	
344.8+03.4 		&	 	Vd 1-3  		&	 	16 49 32.87 		&	 	-39 21 08.90		&	Jun 23, 06	&	 	$2\times1800$	\\	
345.0+03.4 		&	 	Vd 1-4  		&	 	16 50 25.32 		&	 	-39 08 18.90		&	Jun 24, 06	&	 	$3\times900$	\\	
346.2-08.2 		&		IC 4663     		&		17 45 28.37		&		-44 54 15.90		&	Jun 25, 07	&		$2\times900$	\\	
347.7+02.0 		&	 	Vd 1-8  		&	 	17 04 33.77 		&	 	-37 53 14.90		&	Jun 24, 06	&	 	$2\times1200$	\\	
348.0-13.8 		&	 	IC 4699 		&	 	18 18 32.02 		&	 	-45 59 01.70		&	Jun 23, 06	&	 	$3\times900$	\\	
350.5-05.0 		&		  H 1-28 		&		17 42 54.07		&		-39 36 24.00		&	Jun 24, 07	&		$2\times1800$	\\	
350.9+04.4		&		 H 2-1        		&		17 04 36.26		&		-33 59 18.80		&	Jun 21, 07	&		$4\times240$, $1\times120$	\\	
351.6-06.2 		&		  H 1-37 		&		17 50 44.57		&		-39 17 26.00		&	Jun 24, 07	&		$2\times900$	\\	
352.6+03.0		&		  H 1-8  		&		17 14 42.90		&		-33 24 47.20		&	Jun 24, 07	&		$2\times1200$	\\	
355.4-04.0     		&		 Hf 2-1       		&		17 51 12.15		&		-34 55 24.30		&	Jun 25, 07	&		$2\times900$	\\	
355.9+03.6		&		H 1-9			&		17 21 31.90		&		-30 20 48.35		&	Jun 22, 07	&		$3\times500$	\\
356.3-06.2 		&		  M 3-49 		&		18 02 32.11		&		-35 13 14.70		&	Jun 23, 07	&		$2\times1800$	\\	
356.8-05.4		&		  H 2-35 		&		18 00 18.26		&		-34 27 39.30		&	Jun 21, 07	&		$2\times1800$	\\	
357.4-04.6     		&		 M 2-22   		&		17 58 32.63		&		-33 28 36.60		&	Jun 22, 07	&		$2\times1200$	\\	
358.2+03.5		&		H 2-10			&		17 27 32.85		&		-28 31 06.90		&	Jun 21, 07	&		$2\times1200$	\\	
358.3+03.0      	&	        H 1-17          	&	        17 29 40.59     	&	        -28 40 22.10    	&	Jun 23, 07	&	        $3\times900$     \\	
358.7+05.2     		&		 M 3-40       		&		17 22 28.27		&		-27 08 42.40		&	Jun 23, 07	&		$2\times1200$	\\	
358.8+03.0		&		 Th 3-26      		&		17 31 09.30		&		-28 14 50.40		&	Jun 24, 07	&		$2\times1800$	\\	
359.8+03.7     		&		 Th 3-25      		&		17 30 46.72		&		-27 05 59.10		&	Jun 22, 07	&		$2\times1200$	\\	

\end{longtable}

\setcounter{table}{2}

\begin{landscape}

\begin{longtable}{@{\extracolsep{-0.05in}}l c c c c c c c c c c}
\caption{\label{tab_param} Physical parameters for the observed objects. Densities are in units of $10^3$ cm$^{-3}$ and temperatures are in units of $10^4$ K.}\\
\hline\hline
PN G &  E(B-V) & $\sigma_{\mbox{\scriptsize E(B-V)}}$ & $c(H\beta)$ & $\mbox{n}_{\mbox{e}}$([\ion{S}{II}]) &   $\sigma_{\mbox{\scriptsize n}_{\mbox{\scriptsize e}} \mbox{\scriptsize ([\ion{S}{II}])}}$ & T([\ion{N}{II}]) & $\sigma_{\mbox{\scriptsize T ([\ion{N}{II}])}}$   & T([\ion{O}{III}]) & $\sigma_{\mbox{\scriptsize T ([\ion{O}{III}])}}$ & Notes\\
\hline
\endfirsthead
\caption{continued.}\\
\hline\hline
PN G &  E(B-V) & $\sigma_{\mbox{\scriptsize E(B-V)}}$ & $c(H\beta)$ & $\mbox{n}_{\mbox{e}}$([\ion{S}{II}]) &   $\sigma_{\mbox{\scriptsize n}_{\mbox{\scriptsize e}} \mbox{\scriptsize ([\ion{S}{II}])}}$ & T([\ion{N}{II}]) & $\sigma_{\mbox{\scriptsize T ([\ion{N}{II}])}}$   & T([\ion{O}{III}]) & $\sigma_{\mbox{\scriptsize T ([\ion{O}{III}])}}$ & Notes\\
\hline
\endhead
\hline
\endfoot

        000.7-02.7 & 0.23 & 0.04  &  0.33  & 5.84 &   -- & 1.41 & 0.19 & 1.28 & 0.07 &	 \\
	000.9-04.8 & 0.87 & 0.02  &  1.26  & 0.90 & 0.1  & 1.62 & 0.23 & 1.44 & 0.08 &	 \\
	004.0-11.1 & 0.11 & 0.06  &  0.16  & 0.76 & 0.07 & 0.87 & 0.05 & 1.05 & 0.05 &	 \\
	004.2-04.3 & 0.26 &  --   &  0.38  & 5.00 &  --  & 1.07 & 0.06 & 1.07 & 0.06 &	 \\
	005.2-18.6 & 0.17 & 0.03  &  0.25  & 1.90 & 0.1  & 1.10 & 0.10 & 1.23 & 0.05 &	 \\
	005.5-2.5  & 0.86 & 0.02  &  1.24  & 2.20 & 0.2  & 1.25 & 0.10 & 0.99 & 0.05 &	 \\
	006.4-04.6 & 0.12 &  --   &  0.17  & 3.10 & 2.7  & 1.30 & 0.10 & 1.67 & 0.12 & 1 \\
	006.8-03.4 & 1.18 & 0.01  &  1.71  & 6    & 3	 & 1.24 & 0.08 & 1.24 & 0.08 &	 \\
	007.0+06.3 & 0.86 & 0.04  &  1.24  & 5    & 4	 & 1.15 & 0.11 & 1.09 & 0.06 &	 \\
	010.7+07.4 & 0.70 & 0.10  &  1.01  & 0.70 & 0.2  & 2.02 & 0.39 & 1.30 & 0.08 &	 \\
	011.0-05.1 & 0.30 & 0.01  &  0.43  & 3    & 3	 & 1.14 & 0.04 & 1.14 & 0.04 &	 \\
	011.3+02.8 & 1.55 & 0.02  &  2.24  & 1.30 & 0.6  & 1.78 & 0.21 & 1.78 & 0.21 &	 \\
	011.7-06.6 & 0.12 & 0.05  &  0.17  & 10   & 1	 & 0.72 & 0.04 & 0.72 & 0.04 &	 \\
	012.6-02.6 & 1.33 &  --   &  1.92  & 8    & 2	 & 0.67 & 0.04 & 0.67 & 0.04 &	 \\
	013.8-07.9 & 0.38 &  --   &  0.55  & 0.79 & 0.07 & 1.49 & 0.20 & 1.44 & 0.07 &	 \\
	015.9+03.3 & 1.66 & 0.02  &  2.40  & 34   & 30   & 0.76 & 0.04 & 0.76 & 0.04 &	 \\
	016.4-01.9 & 0.19 & 0.06  &  0.27  & 2.00 & 0.4  & 0.81 & 0.04 & 2.16 & 0.22 &	 \\
	019.7-04.5 & 1.21 & 0.01  &  1.75  & 6.90 & 0.8  & 1.16 & 0.07 & 1.06 & 0.04 &	 \\
	021.8-00.4 & 1.21 & 0.04  &  1.75  & 1.90 & 0.3  & 1.06 & 0.07 & 1.32 & 0.10 &	 \\
	023.0+04.3 & 1.21 & 0.05  &  1.75  & 1.10 & 0.7  & 2.45 & 0.56 & 1.18 & 0.09 &	 \\
	023.3-07.6 & 0.20 & 0.03  &  0.29  & 0.57 & 0.02 & 0.90 & 0.05 & 1.06 & 0.05 &	 \\
	023.8-01.7 & 1.89 & 0.06  &  2.73  & 9    & 6	 & 0.81 & 0.06 & 0.81 & 0.06 &	 \\
	024.1+03.8 & 0.80 & 0.60  &  1.16  & 5.04 &   -- & 0.86 & 0.13 & 1.16 & 0.17 &	 \\
	025.9-02.1 & 0.89 & 0.03  &  1.29  & 1.60 & 0.9  & 2.17 & 0.41 & 0.90 & 0.04 &	 \\
	335.4-01.1 & 2.00 & 0.10  &  2.89  & 1.24 & 0.03 & 1.24 & 0.10 & 2.49 & 0.39 &	 \\
	335.9-03.6 & 0.92 &  --   &  1.33  & 0.20 & 0.2  & 1.40 & 0.60 & 1.47 & 0.12 & 1 \\
	336.2+01.9 & 1.50 & 0.50  &  2.17  & 1.00 & 0.1  & 2.82 & 1.17 & 1.51 & 0.26 &	 \\
	336.3-05.6 & 0.52 & 0.05  &  0.75  & 3.52 & 0.03 & 1.10 & 0.07 & 1.33 & 0.06 &	 \\
	336.9+08.3 & 0.66 & 0.07  &  0.95  & 5    &   -- & 1.23 & 0.06 & 1.23 & 0.06 &	 \\
	338.8+05.6 & 0.56 & 0.02  &  0.81  & 1.20 & 0.3  & 1.01 & 0.06 & 1.01 & 0.04 &	 \\
	340.9-04.6 & 0.95 & 0.01  &  1.37  & 3.60 & 0.3  & 1.45 & 0.17 & 1.18 & 0.06 &	 \\
	342.9-04.9 & 0.29 & 0.01  &  0.42  & 0.53 & 0.05 & 1.02 & 0.06 & 1.30 & 0.06 &	 \\
	343.0-01.7 & 1.92 &  --   &  2.78  & 5.5  & 0.3  & 1.72 & 0.27 & 1.13 & 0.08 &	 \\
	344.2-01.2 & 1.00 & 0.10  &  1.45  & 0.80 & 0.2  & 1.03 & 0.09 & 1.74 & 0.19 &	 \\
	344.4+02.8 & 1.07 & 0.01  &  1.55  & 0.8  & 0.3  & 1.74 & 0.30 & 1.28 & 0.07 &	 \\
	344.8+03.4 & 0.82 & 0.06  &  1.19  & 0.71 & 0.08 & 0.81 & 0.05 & 0.81 & 0.05 &	 \\
	345.0+03.4 & 0.83 & 0.07  &  1.20  & 7.50 & 0.9  & 1.48 & 0.16 & 1.34 & 0.07 &	 \\
	346.2-08.2 & 0.40 & 0.10  &  0.58  & 3.81 &   -- & 1.28 & 0.11 & 1.15 & 0.06 &	 \\
	347.7+02.0 & 1.84 & 0.01  &  2.66  & 10   & 2	 & 2.00 & 0.32 & 1.88 & 0.15 &	 \\
	348.0-13.8 & 0.20 & 0.20  &  0.29  & 1.39 &   -  & 1.27 & 0.07 & 1.27 & 0.07 &	 \\
	350.5-05.0 & 0.67 & 0.02  &  0.97  & 0.70 & 0.2  & 0.86 & 0.04 & 1.19 & 0.06 &	 \\
	350.9+04.4 & 0.49 & 0.05  &  0.71  & 5    & 3	 & 1.30 & 0.07 & 1.83 & 0.11 &	 \\
	351.6-06.2 & 0.48 & 0.03  &  0.69  & 1.22 &   -- & 1.01 & 0.06 & 1.28 & 0.06 &	 \\
	352.6+03.0 & 1.42 &  --   &  2.05  & 7.00 & 0.8  & 0.74 & 0.05 & 1.40 & 0.11 &	 \\
	355.4-04.0 & 0.53 & 0.01  &  0.77  & 0.6  & 0.1  & 1.29 & 0.11 & 1.28 & 0.06 &	 \\
	355.9+03.6 & 1.00 & 0.10  &  1.45  & 13   & 9	 & 1.94 & 0.28 & 1.58 & 0.13 &	 \\
	356.3-06.2 & 0.35 & 0.06  &  0.51  & 0.24 & 0.03 & 0.90 & 0.06 & 1.33 & 0.09 &	 \\
	356.8-05.4 & 0.63 & 0.02  &  0.91  & 0.24 & 0.06 & 0.91 & 0.07 & 1.19 & 0.07 &	 \\
	357.4-04.6 & 0.69 & 0.05  &  1.00  & 1.69 &   -- & 1.23 & 0.10 & 1.22 & 0.07 &	 \\
	358.2+03.5 & 1.57 & 0.01  &  2.27  & 7    & 3	 & 1.74 & 0.24 & 1.28 & 0.07 &	 \\
	358.3+03.0 & 1.39 & 0.07  &  2.01  & 19   & 6	 & 1.98 & 0.29 & 1.40 & 0.09 &	 \\
	358.7+05.2 & 1.53 & 0.04  &  2.21  & 8.99 & 0.08 & 0.68 & 0.04 & 0.68 & 0.04 &	 \\
	358.8+03.0 & 1.12 &  --   &  1.62  & 2    & 2	 & 1.31 & 0.14 & 1.58 & 0.13 &	 \\
	359.8+03.7 & 1.60 & 0.06  &  2.31  & 5.21 & 0.08 & 1.31 & 0.13 & 1.55 & 0.14 &	 \\

\multicolumn{8}{l}{\footnotesize{1 T([\ion{N}{II}]) and $\sigma_{\mbox{\scriptsize T ([\ion{N}{II}])}}$ were obtained from the mean of individual measure for each object.}}\\

\end{longtable}

\end{landscape}

\begin{landscape}
\setlength{\tabcolsep}{3pt}
\small
\begin{longtable}{l c c c c c c c c c c c c c c c c c c} 
\caption{\label{tab_ionic_abund} Ionic abundances relative to hydrogen. }\\
\hline\hline
PN G  & He$^+$  & $\sigma_{\mbox{He}^+}$ & He$^{++}$  & $\sigma_{\mbox{He}^{++}}$ & N$^+$  	    & $\sigma_{\mbox{N}^+}$ & S$^+$           & $\sigma_{\mbox{S}^+}$ & S$^{++}$           & $\sigma_{\mbox{S}^{++}}$ & O$^+$            & $\sigma_{\mbox{O}^+}$ & O$^{++}$           & $\sigma_{\mbox{O}^{++}}$ & Ar$^{++}$          & $\sigma_{\mbox{Ar}^{++}}$ & Ne$^{++}$          & $\sigma_{\mbox{Ne}^{++}}$\\
      & 		 &			      &			  &				& $\phantom{}^{\times10^6}$ & $\phantom{}^{\times10^6}$ & $\phantom{}^{\times10^7}$ & $\phantom{}^{\times10^7}$ & $\phantom{}^{\times10^6}$ & $\phantom{}^{\times10^6}$  & $\phantom{}^{\times10^5}$ & $\phantom{}^{\times10^5}$ & $\phantom{}^{\times10^4}$ & $\phantom{}^{\times10^4}$  & $\phantom{}^{\times10^6}$ & $\phantom{}^{\times10^6}$   & $\phantom{}^{\times10^5}$ & $\phantom{}^{\times10^5}$  \\		
\hline
\endfirsthead
\caption{continued.}\\
\hline\hline

PN G  & He$^+$  & $\sigma_{\mbox{He}^+}$ & He$^{++}$  & $\sigma_{\mbox{He}^{++}}$ & N$^+$  	    & $\sigma_{\mbox{N}^+}$ & S$^+$           & $\sigma_{\mbox{S}^+}$ & S$^{++}$           & $\sigma_{\mbox{S}^{++}}$ & O$^+$            & $\sigma_{\mbox{O}^+}$ & O$^{++}$           & $\sigma_{\mbox{O}^{++}}$ & Ar$^{++}$          & $\sigma_{\mbox{Ar}^{++}}$ & Ne$^{++}$          & $\sigma_{\mbox{Ne}^{++}}$\\
      & 		 &			      &			  &				& $\phantom{}^{\times10^6}$ & $\phantom{}^{\times10^6}$ & $\phantom{}^{\times10^7}$ & $\phantom{}^{\times10^7}$ & $\phantom{}^{\times10^6}$ & $\phantom{}^{\times10^6}$  & $\phantom{}^{\times10^5}$ & $\phantom{}^{\times10^5}$ & $\phantom{}^{\times10^4}$ & $\phantom{}^{\times10^4}$  & $\phantom{}^{\times10^6}$ & $\phantom{}^{\times10^6}$   & $\phantom{}^{\times10^5}$ & $\phantom{}^{\times10^5}$  \\		
\hline
\endhead
\hline
\endfoot

	000.7-02.7			&	0.090	&	0.016	&	0.020	&	0.003	&	2.08	&	0.63	&	0.86	&	0.47	&	3.23	&	0.89	&	1.73	&	1.04	&	2.25	&	0.44	&	0.47	&	0.10	&	1.80	&	0.42	\\
	000.9-04.8			&	0.029	&	0.005	&	0.089	&	0.011	&	0.87	&	0.26	&	0.73	&	0.18	&	2.30	&	0.55	&	0.28	&	0.16	&	1.45	&	0.27	&	1.02	&	0.17	&	3.31	&	0.69	\\
	004.0-11.1 			&	0.088	&	0.012	&	--	&	--	&	13.73	&	3.64	&	2.67	&	0.74	&	1.22	&	0.33	&	10.85	&	5.92	&	1.72	&	0.30	&	0.80	&	0.16	&	7.40	&	1.45	\\
	004.2-04.3      		&	0.101	&	0.020	&	0.005	&	--	&	0.75	&	0.16	&	0.46	&	0.08	&	1.01	&	0.36	&	2.17	&	0.26	&	2.85	&	0.85	&	0.66	&	0.10	&	5.48	&	1.34	\\
	005.2-18.6 			&	0.073	&	0.010	&	0.034	&	0.004	&	3.66	&	0.94	&	1.49	&	0.53	&	1.74	&	0.40	&	2.43	&	1.31	&	2.05	&	0.35	&	0.74	&	0.12	&	3.72	&	0.68	\\
	005.5-2.5			&	0.132	&	0.019	&	0.018	&	0.001	&	8.80	&	1.97	&	2.78	&	0.79	&	2.64	&	0.75	&	1.61	&	0.69	&	2.50	&	0.54	&	1.72	&	0.31	&	7.39	&	1.78	\\
	006.4-04.6 			&	0.056	&	0.020	&	0.086	&	0.014	&	0.48	&	0.14	&	0.97	&	0.28	&	2.82	&	0.76	&	0.19	&	0.07	&	1.01	&	0.22	&	1.16	&	0.24	&	1.52	&	0.37	\\
	006.8-03.4 			&	0.096	&	0.014	&	--	&	--	&	0.71	&	0.13	&	0.30	&	0.12	&	0.79	&	0.22	&	0.51	&	0.21	&	1.82	&	0.37	&	0.30	&	0.06	&	3.57	&	0.88	\\
	007.0+06.3 			&	0.056	&	0.008	&	--	&	--	&	5.03	&	1.40	&	1.34	&	0.58	&	1.33	&	0.38	&	1.47	&	0.84	&	1.32	&	0.30	&	1.21	&	0.25	&	3.23	&	0.78	\\
	010.7+07.4			&	0.056	&	0.013	&	0.096	&	0.013	&	0.56	&	0.24	&	0.51	&	0.19	&	2.88	&	0.99	&	0.23	&	0.16	&	1.47	&	0.31	&	1.08	&	0.33	&	3.32	&	0.74	\\
	011.0-05.1			&	0.095	&	0.011	&	0.003	&	--	&	--	&	--	&	0.14	&	0.03	&	0.78	&	0.16	&	0.51	&	0.19	&	2.81	&	0.42	&	0.38	&	0.05	&	5.50	&	0.92	\\
	011.3+02.8 			&	0.111	&	0.018	&	0.007	&	0.001	&	0.62	&	0.14	&	0.14	&	0.04	&	0.21	&	0.07	&	0.07	&	0.03	&	0.07	&	0.02	&	0.04	&	0.01	&	0.88	&	0.26	\\
	011.7-06.6 			&	0.005	&	0.001	&	--	&	--	&	100.14	&	25.83	&	41.91	&	27.63	&	--	&	--	&	42.87	&	23.47	&	0.01	&	0.00	&	0.12	&	0.04	&	--	&	--	\\
	012.6-02.6			&	0.015	&	0.003	&	--	&	--	&	144.76	&	35.43	&	34.63	&	15.82	&	3.69	&	1.50	&	26.93	&	13.69	&	0.05	&	0.02	&	0.22	&	0.06	&	--	&	--	\\
	013.8-07.9 			&	0.044	&	0.007	&	0.086	&	0.009	&	0.64	&	0.19	&	0.40	&	0.11	&	1.27	&	0.29	&	0.63	&	0.36	&	0.95	&	0.15	&	0.54	&	0.09	&	1.79	&	0.33	\\
	015.9+03.3 			&	0.087	&	0.012	&	0.002	&	--	&	136.70	&	37.17	&	27.17	&	33.28	&	8.27	&	5.20	&	41.15	&	27.18	&	0.66	&	0.35	&	2.42	&	1.04	&	--	&	--	\\
	016.4-01.9			&	0.100	&	0.013	&	0.005	&	--	&	54.01	&	13.02	&	13.36	&	3.71	&	0.46	&	0.13	&	88.74	&	41.64	&	0.04	&	0.01	&	0.49	&	0.10	&	0.02	&	0.00	\\
	019.7-04.5			&	0.129	&	0.013	&	--	&	--	&	19.44	&	3.60	&	3.67	&	1.57	&	3.94	&	0.74	&	2.55	&	0.83	&	3.05	&	0.49	&	1.65	&	0.21	&	9.32	&	1.68	\\
	021.8-00.4     			&	0.122	&	0.021	&	0.015	&	0.002	&	89.34	&	21.72	&	4.20	&	1.36	&	1.32	&	0.41	&	16.94	&	8.06	&	1.99	&	0.45	&	1.78	&	0.38	&	3.53	&	0.90	\\
	023.0+04.3 			&	0.069	&	0.022	&	--	&	--	&	0.26	&	0.10	&	0.09	&	0.03	&	0.82	&	0.27	&	0.07	&	0.05	&	1.18	&	0.31	&	0.62	&	0.14	&	3.21	&	0.95	\\
	023.3-07.6 			&	0.157	&	0.022	&	0.014	&	0.002	&	170.57	&	36.57	&	41.89	&	8.49	&	4.02	&	1.11	&	54.63	&	22.50	&	2.37	&	0.47	&	2.24	&	0.46	&	7.73	&	1.74	\\
	023.8-01.7     			&	0.007	&	0.001	&	--	&	--	&	63.48	&	17.88	&	22.68	&	11.82	&	1.90	&	0.92	&	14.89	&	8.30	&	0.03	&	0.01	&	--	&	--	&	15.51	&	6.72	\\
	024.1+03.8 			&	0.143	&	0.089	&	--	&	--	&	68.34	&	91.73	&	6.66	&	9.81	&	1.56	&	2.07	&	32.99	&	72.20	&	0.56	&	0.32	&	2.24	&	3.13	&	--	&	--	\\
	025.9-02.1  			&	0.102	&	0.024	&	0.005	&	0.001	&	0.51	&	0.16	&	0.14	&	0.05	&	2.32	&	0.68	&	0.12	&	0.07	&	3.93	&	0.89	&	1.69	&	0.33	&	8.26	&	2.07	\\
	335.4-01.1			&	0.128	&	0.030	&	0.131	&	0.026	&	146.10	&	47.89	&	23.99	&	8.04	&	1.53	&	0.64	&	7.88	&	4.91	&	0.50	&	0.15	&	1.05	&	0.36	&	2.11	&	0.67	\\
	335.9-03.6 			&	0.055	&	0.061	&	0.051	&	0.008	&	0.12	&	0.04	&	0.09	&	0.04	&	0.42	&	0.14	&	0.08	&	0.05	&	0.74	&	0.17	&	0.48	&	0.10	&	0.44	&	0.13	\\
	336.2+01.9 			&	0.034	&	0.029	&	0.028	&	0.006	&	0.40	&	0.45	&	0.19	&	0.23	&	0.26	&	0.29	&	0.05	&	0.09	&	0.86	&	0.41	&	0.60	&	0.68	&	1.57	&	0.44	\\
	336.3-05.6     			&	0.087	&	0.012	&	0.049	&	0.005	&	39.51	&	7.94	&	11.04	&	3.44	&	3.34	&	0.65	&	12.79	&	4.67	&	2.45	&	0.36	&	1.35	&	0.22	&	4.46	&	0.73	\\
	336.9+08.3 			&	0.096	&	0.014	&	--	&	--	&	--	&	--	&	--	&	--	&	0.70	&	0.20	&	--	&	--	&	1.64	&	0.29	&	0.44	&	0.10	&	3.29	&	0.66	\\
	338.8+05.6			&	0.125	&	0.013	&	0.005	&	--	&	8.66	&	1.77	&	2.43	&	0.48	&	2.69	&	0.55	&	3.65	&	1.50	&	3.56	&	0.53	&	1.10	&	0.15	&	8.71	&	1.67	\\
	340.9-04.6 			&	0.102	&	0.018	&	0.002	&	--	&	1.06	&	0.29	&	0.64	&	0.26	&	1.63	&	0.40	&	0.55	&	0.26	&	2.45	&	0.45	&	0.64	&	0.11	&	4.87	&	1.13	\\
	342.9-04.9 			&	0.083	&	0.012	&	0.056	&	0.007	&	71.14	&	14.06	&	22.87	&	4.24	&	6.26	&	1.39	&	16.66	&	6.88	&	2.50	&	0.39	&	2.55	&	0.35	&	4.04	&	0.70	\\
	343.0-01.7 			&	0.100	&	0.021	&	0.004	&	0.001	&	2.41	&	0.73	&	0.63	&	0.31	&	1.69	&	0.58	&	0.69	&	0.42	&	2.10	&	0.60	&	0.96	&	0.21	&	6.71	&	2.43	\\
	344.2-01.2			&	0.202	&	0.043	&	0.022	&	0.005	&	137.33	&	53.79	&	32.06	&	12.37	&	1.71	&	0.76	&	22.33	&	15.77	&	0.64	&	0.18	&	1.89	&	0.73	&	1.80	&	0.58	\\
	344.4+02.8 			&	0.072	&	0.016	&	0.033	&	0.005	&	0.31	&	0.10	&	0.15	&	0.06	&	0.70	&	0.19	&	0.51	&	0.34	&	2.29	&	0.46	&	0.56	&	0.11	&	2.93	&	0.69	\\
	344.8+03.4 			&	0.094	&	0.016	&	--	&	--	&	49.22	&	14.96	&	20.62	&	5.91	&	--	&	--	&	41.18	&	24.15	&	0.88	&	0.27	&	1.33	&	0.35	&	--	&	--	\\
	345.0+03.4 			&	0.104	&	0.018	&	0.004	&	--	&	4.26	&	1.30	&	0.92	&	0.48	&	1.06	&	0.29	&	1.89	&	0.96	&	1.74	&	0.32	&	0.55	&	0.11	&	3.89	&	0.76	\\
	346.2-08.2 			&	0.042	&	0.008	&	0.074	&	0.008	&	3.94	&	1.40	&	1.91	&	0.70	&	3.39	&	1.19	&	0.89	&	0.55	&	2.39	&	0.44	&	1.93	&	0.65	&	6.17	&	1.05	\\
	347.7+02.0 			&	0.085	&	0.019	&	0.013	&	0.003	&	1.44	&	0.39	&	0.77	&	0.37	&	0.62	&	0.16	&	0.37	&	0.17	&	1.13	&	0.23	&	0.39	&	0.07	&	2.27	&	0.58	\\
	348.0-13.8 			&	0.055	&	0.011	&	0.025	&	0.003	&	--	&	--	&	0.05	&	0.02	&	0.55	&	0.21	&	0.66	&	0.41	&	2.09	&	0.40	&	0.21	&	0.08	&	4.04	&	0.69	\\
	350.5-05.0 			&	0.169	&	0.025	&	0.013	&	0.002	&	124.70	&	25.65	&	24.32	&	4.61	&	1.31	&	0.39	&	30.46	&	12.83	&	0.98	&	0.20	&	1.40	&	0.25	&	3.38	&	0.77	\\
	350.9+04.4			&	0.059	&	0.006	&	--	&	--	&	9.27	&	1.60	&	1.26	&	0.39	&	0.49	&	0.10	&	7.69	&	2.34	&	0.06	&	0.01	&	0.18	&	0.03	&	0.09	&	0.02	\\
	351.6-06.2 			&	0.084	&	0.012	&	0.069	&	0.008	&	43.55	&	9.86	&	11.97	&	2.84	&	3.33	&	0.75	&	5.20	&	2.27	&	1.35	&	0.25	&	1.47	&	0.24	&	4.21	&	0.85	\\
	352.6+03.0			&	0.050	&	0.008	&	0.004	&	0.001	&	125.01	&	31.92	&	23.51	&	10.80	&	1.09	&	0.32	&	58.87	&	28.77	&	0.82	&	0.20	&	1.62	&	0.30	&	2.20	&	0.59	\\
	355.4-04.0     			&	0.047	&	0.007	&	0.081	&	0.010	&	8.15	&	1.75	&	3.68	&	0.77	&	3.53	&	0.72	&	1.17	&	0.50	&	2.31	&	0.38	&	1.93	&	0.29	&	4.34	&	0.74	\\
	355.9+03.6			&	0.063	&	0.014	&	--	&	--	&	4.63	&	1.80	&	0.35	&	0.23	&	1.19	&	0.44	&	3.75	&	2.53	&	0.16	&	0.03	&	0.34	&	0.13	&	0.16	&	0.04	\\
	356.3-06.2 			&	0.100	&	0.018	&	0.019	&	0.004	&	93.51		24.10	&	23.92	&	6.03	&	--	&	--	&	16.03	&	8.94	&	0.82	&	0.19	&	1.03	&	0.24	&	1.65	&	0.47	\\
	356.8-05.4			&	0.074	&	0.012	&	0.023	&	0.004	&	29.19	&	7.79	&	9.29	&	2.34	&	1.05	&	0.32	&	--	&	--	&	1.56	&	0.35	&	1.29	&	0.27	&	3.95	&	0.97	\\
	357.4-04.6     			&	0.089	&	0.015	&	0.028	&	0.004	&	20.69	&	5.22	&	5.26	&	1.64	&	1.29	&	0.39	&	1.80	&	0.84	&	1.51	&	0.32	&	1.19	&	0.24	&	4.01	&	0.98	\\
	358.2+03.5			&	0.090	&	0.014	&	--	&	--	&	0.85	&	0.23	&	0.39	&	0.18	&	0.92	&	0.22	&	0.27	&	0.13	&	1.67	&	0.31	&	0.42	&	0.07	&	6.01	&	1.36	\\
	358.3+03.0      		&	0.107	&	0.020	&	0.006	&	0.001	&	5.35	&	1.55	&	1.32	&	0.79	&	3.13	&	0.86	&	0.98	&	0.48	&	2.06	&	0.44	&	1.46	&	0.35	&	3.22	&	0.78	\\
	358.7+05.2     			&	0.020	&	0.004	&	--	&	--	&	137.87	&	43.56	&	43.95	&	20.69	&	10.30	&	4.89	&	33.52	&	20.19	&	--	&	--	&	0.25	&	0.09	&	48.77	&	21.48	\\
	358.8+03.0			&	0.046	&	0.010	&	0.080	&	0.014	&	11.38	&	3.15	&	4.08	&	1.37	&	3.53	&	1.07	&	1.61	&	0.93	&	1.00	&	0.24	&	0.75	&	0.15	&	2.62	&	0.74	\\
	359.8+03.7     			&	0.097	&	0.018	&	--	&	--	&	4.32	&	1.28	&	1.01	&	0.54	&	0.52	&	0.17	&	2.09	&	1.13	&	0.49	&	0.12	&	0.34	&	0.09	&	1.01	&	0.29	\\

\end{longtable}
\end{landscape}

\begin{small}

\begin{longtable}{@{\extracolsep{-0.1in}} l c c c c c c c c c c c c}
\caption{\label{tab_abund} New chemical abundances and errors in the usual notation.}\\
\hline\hline
PN G & He/H & $\sigma_{{\mbox{\scriptsize He/H}}}$ & $\epsilon(\mbox{N})$ & $\sigma_{\epsilon(\mbox{\scriptsize N})}$ &  $\epsilon(\mbox{S})$& $\sigma_{\epsilon(\mbox{\scriptsize S})}$ & $\epsilon(\mbox{O})$ & $\sigma_{\epsilon(\mbox{\scriptsize O})}$ & $\epsilon(\mbox{Ar})$ & $\sigma_{\epsilon(\mbox{\scriptsize Ar})}$ & $\epsilon(\mbox{Ne})$ & $\sigma_{\epsilon(\mbox{\scriptsize Ne})}$ \\
\hline
\endfirsthead
\caption{continued.}\\
\hline\hline
PN G & He/H & $\sigma_{{\mbox{\scriptsize He/H}}}$ & $\epsilon(\mbox{N})$ & $\sigma_{\epsilon(\mbox{\scriptsize N})}$ &  $\epsilon(\mbox{S})$& $\sigma_{\epsilon(\mbox{\scriptsize S})}$ & $\epsilon(\mbox{O})$ & $\sigma_{\epsilon(\mbox{\scriptsize O})}$ & $\epsilon(\mbox{Ar})$ & $\sigma_{\epsilon(\mbox{\scriptsize Ar})}$ & $\epsilon(\mbox{Ne})$ & $\sigma_{\epsilon(\mbox{\scriptsize Ne})}$ \\
\hline
\endhead
\hline
\endfoot

	000.7-02.7			&	0.110	&	0.017	&	7.55	&	0.15	&	6.79	&	0.14	&	8.47	&	0.08	&	5.92	&	0.09	&	7.38	&	0.10	\\
	000.9-04.8			&	0.118	&	0.013	&	8.27	&	0.18	&	6.99	&	0.14	&	8.79	&	0.10	&	6.76	&	0.09	&	8.14	&	0.10	\\
	004.0-11.1 			&	0.088	&	0.012	&	7.56	&	0.08	&	6.22	&	0.11	&	8.45	&	0.11	&	6.24	&	0.14	&	8.08	&	0.11	\\
	004.2-04.3      		&	0.106	&	0.021	&	8.02*	&	--	&	6.66*	&	--	&	8.45*	&	--	&	5.99	&	0.06	&	7.79	&	0.10	\\
	005.2-18.6 			&	0.106	&	0.012	&	7.71	&	0.13	&	6.51	&	0.13	&	8.54	&	0.07	&	6.22	&	0.08	&	7.79	&	0.08	\\
	005.5-2.5			&	0.149	&	0.021	&	8.22	&	0.14	&	6.74	&	0.14	&	8.48	&	0.09	&	6.45	&	0.08	&	7.96	&	0.10	\\
	006.4-04.6 			&	0.145	&	0.028	&	7.80	&	0.20	&	7.00	&	0.16	&	8.40	&	0.13	&	6.59	&	0.13	&	7.58	&	0.14	\\
	006.8-03.4 			&	0.096	&	0.014	&	7.40	&	0.12	&	6.28	&	0.12	&	8.27	&	0.09	&	5.62	&	0.08	&	7.57	&	0.11	\\
	007.0+06.3 			&	0.051	&	0.012	&	7.70	&	0.13	&	6.37	&	0.14	&	8.17	&	0.09	&	6.26	&	0.09	&	7.56	&	0.11	\\
	010.7+07.4			&	0.151	&	0.020	&	8.02	&	0.20	&	7.07	&	0.16	&	8.62	&	0.10	&	6.61	&	0.13	&	7.97	&	0.13	\\
	011.0-05.1			&	0.098	&	0.012	&	--	&	--	&	6.34	&	0.09	&	8.47	&	0.07	&	5.73	&	0.07	&	7.76	&	0.07	\\
	011.3+02.8 			&	0.118	&	0.019	&	6.83	&	0.10	&	5.55	&	0.13	&	6.89	&	0.11	&	4.83	&	0.09	&	7.02	&	0.13	\\
	011.7-06.6 $\star$		&	--	&	--	&	8.00	&	0.11	&	7.38	&	0.28	&	8.63	&	0.23	&	7.74	&	0.22	&	--	&	--	\\
	012.6-02.6 $\star$		&	--	&	--	&	8.23	&	0.10	&	6.86	&	0.18	&	8.51	&	0.22	&	7.28	&	0.22	&	--	&	--	\\
	013.8-07.9 			&	0.131	&	0.013	&	7.46	&	0.17	&	6.51	&	0.13	&	8.48	&	0.08	&	6.37	&	0.08	&	7.76	&	0.09	\\
	015.9+03.3 $\star$		&	0.089	&	0.012	&	8.21	&	0.10	&	7.03	&	0.31	&	8.70	&	0.29	&	7.38	&	0.24	&	--	&	--	\\
	016.4-01.9 $\star$		&	0.104	&	0.013	&	7.75	&	0.10	&	6.26	&	0.10	&	8.97	&	0.20	&	8.15	&	0.22	&	7.62	&	0.21	\\
	019.7-04.5			&	0.129	&	0.013	&	8.39	&	0.09	&	6.86	&	0.09	&	8.52	&	0.07	&	6.38	&	0.06	&	8.01	&	0.09	\\
	021.8-00.4 $\star$		&	0.138	&	0.023	&	8.35	&	0.09	&	6.29	&	0.12	&	8.63	&	0.12	&	6.70	&	0.12	&	7.87	&	0.12	\\
	023.0+04.3 			&	0.063	&	0.028	&	7.62	&	0.20	&	6.50	&	0.19	&	8.08	&	0.11	&	5.92	&	0.10	&	7.51	&	0.12	\\
	023.3-07.6 $\star$ 		&	0.171	&	0.022	&	8.43	&	0.06	&	6.92	&	0.08	&	8.93	&	0.13	&	7.03	&	0.14	&	8.45	&	0.13	\\
	023.8-01.7 $\star$		&	--	&	--	&	7.92	&	0.12	&	6.62	&	0.21	&	8.29	&	0.24	&	--	&	--	&	9.97	&	0.26	\\
	024.1+03.8 $\star$		&	0.136	&	0.098	&	7.88	&	0.43	&	6.37	&	0.59	&	8.55	&	0.85	&	7.28	&	1.22	&	--	&	--	\\
	025.9-02.1  			&	0.107	&	0.025	&	8.25	&	0.17	&	7.06	&	0.17	&	8.62	&	0.10	&	6.38	&	0.08	&	7.94	&	0.11	\\
	335.4-01.1			&	0.261	&	0.044	&	8.70	&	0.10	&	6.66	&	0.13	&	8.43	&	0.17	&	6.88	&	0.22	&	8.05	&	0.16	\\
	335.9-03.6 			&	0.118	&	0.050	&	7.29	&	0.30	&	6.21	&	0.21	&	8.13	&	0.16	&	6.07	&	0.16	&	6.91	&	0.18	\\
	336.2+01.9 			&	0.165*	&	0.014	&	8.07	&	0.27	&	6.12	&	0.29	&	8.20	&	0.12	&	6.15	&	0.34	&	7.43	&	0.20	\\
	336.3-05.6     			&	0.136	&	0.015	&	8.26	&	0.07	&	6.75	&	0.09	&	8.77	&	0.08	&	6.64	&	0.09	&	8.03	&	0.08	\\
	336.9+08.3 			&	0.096	&	0.014	&	--	&	--	&	5.84	&	0.13	&	8.22	&	0.08	&	5.77	&	0.10	&	7.52	&	0.09	\\
	338.8+05.6			&	0.130	&	0.013	&	7.98	&	0.11	&	6.67	&	0.10	&	8.61	&	0.06	&	6.23	&	0.06	&	8.00	&	0.08	\\
	340.9-04.6 			&	0.104	&	0.017	&	7.69	&	0.14	&	6.63	&	0.14	&	8.41	&	0.08	&	5.96	&	0.08	&	7.70	&	0.10	\\
	342.9-04.9 			&	0.139	&	0.016	&	8.49	&	0.07	&	7.02	&	0.08	&	8.85	&	0.08	&	6.99	&	0.09	&	8.06	&	0.09	\\
	343.0-01.7 			&	0.103	&	0.022	&	7.88	&	0.19	&	6.60	&	0.18	&	8.36	&	0.12	&	6.14	&	0.09	&	7.86	&	0.16	\\
	344.2-01.2 $\star$		&	0.225	&	0.046	&	8.30	&	0.12	&	6.70	&	0.16	&	8.51	&	0.24	&	7.10	&	0.32	&	7.95	&	0.21	\\
	344.4+02.8 			&	0.106	&	0.017	&	7.31	&	0.19	&	6.30	&	0.16	&	8.54	&	0.09	&	6.05	&	0.08	&	7.64	&	0.11	\\
	344.8+03.4 $\star$      	&	0.094	&	0.016	&	7.78	&	0.12	&	7.11	&	0.12	&	8.70	&	0.23	&	7.00	&	0.22	&	--	&	--	\\
	345.0+03.4 			&	0.107	&	0.019	&	7.64	&	0.13	&	6.27	&	0.13	&	8.30	&	0.08	&	5.93	&	0.10	&	7.65	&	0.08	\\
	346.2-08.2 			&	0.118	&	0.012	&	8.49	&	0.16	&	7.03	&	0.12	&	8.84	&	0.07	&	6.87	&	0.12	&	8.25	&	0.09	\\
	347.7+02.0 			&	0.099	&	0.020	&	7.70	&	0.14	&	6.23	&	0.15	&	8.13	&	0.09	&	5.80	&	0.09	&	7.44	&	0.11	\\
	348.0-13.8 			&	0.079	&	0.012	&	--	&	--	&	6.15	&	0.13	&	8.50	&	0.08	&	5.62	&	0.15	&	7.78	&	0.08	\\
	350.5-05.0 $\star$		&	0.183	&	0.026	&	8.26	&	0.06	&	6.58	&	0.07	&	8.64	&	0.14	&	6.92	&	0.14	&	8.19	&	0.14	\\
	350.9+04.4 $\star$		&	0.059	&	0.007	&	7.00	&	0.07	&	5.79	&	0.09	&	7.92	&	0.12	&	6.52	&	0.15	&	7.08	&	0.13	\\
	351.6-06.2 			&	0.154	&	0.017	&	8.46	&	0.09	&	6.80	&	0.09	&	8.54	&	0.08	&	6.70	&	0.09	&	8.04	&	0.10	\\
	352.6+03.0 $\star$		&	0.073*	&	--	&	8.19	&	0.09	&	6.55	&	0.15	&	8.86	&	0.19	&	7.28	&	0.18	&	8.29	&	0.20	\\
	355.4-04.0     			&	0.127	&	0.014	&	8.66	&	0.14	&	7.02	&	0.11	&	8.83	&	0.09	&	6.88	&	0.08	&	8.10	&	0.09	\\
	355.9+03.6 $\star$		&	0.063	&	0.014	&	6.82	&	0.10	&	6.11	&	0.16	&	7.73	&	0.23	&	6.19	&	0.31	&	6.71	&	0.18	\\
	356.3-06.2 $\star$		&	0.119	&	0.019	&	8.24	&	0.08	&	7.20	&	0.09	&	8.46	&	0.17	&	6.68	&	0.18	&	7.77	&	0.18	\\
	356.8-05.4			&	0.098	&	0.013	&	--	&	--	&	6.31	&	0.10	&	8.31	&	0.10	&	6.36	&	0.09	&	7.72	&	0.11	\\
	357.4-04.6     			&	0.117	&	0.016	&	8.42	&	0.13	&	6.49	&	0.12	&	8.35	&	0.09	&	6.37	&	0.09	&	7.78	&	0.10	\\
	358.2+03.5			&	0.091	&	0.015	&	7.72	&	0.14	&	6.42	&	0.14	&	8.24	&	0.09	&	5.77	&	0.08	&	7.79	&	0.10	\\
	358.3+03.0      		&	0.113	&	0.021	&	8.08	&	0.12	&	6.82	&	0.13	&	8.36	&	0.09	&	6.34	&	0.11	&	7.55	&	0.11	\\
	358.7+05.2 $\star$		&	0.020*	&	--	&	8.14	&	0.14	&	7.17	&	0.20	&	8.52	&	0.26	&	--	&	--	&	--	&	--	\\
	358.8+03.0			&	0.126	&	0.019	&	8.35	&	0.17	&	6.88	&	0.16	&	8.51	&	0.11	&	6.51	&	0.10	&	7.92	&	0.12	\\
	359.8+03.7     			&	0.097	&	0.018	&	7.14	&	0.10	&	5.87	&	0.14	&	7.86	&	0.11	&	5.83	&	0.13	&	7.17	&	0.12	\\

\end{longtable}
\vspace{-0.5cm}
{\footnotesize{* Abundances were calculated from the mean of each measure of the same object.}}

{\footnotesize{$\star$ There is a substantial contribution of neutral helium not taken into account.}}
\end{small}


\begin{thebibliography}

 \bibitem[Acker et al. (1992)]{acker92} Acker, A., Marcout, J., Ochsenbein, F., Stenholm, B., \& Tylenda, R.\ 1992, (Garching: ESO) 

  \bibitem[Alexander \& Balick (1997)]{alex97} Alexander, J., \& Balick, B.\ 1997, \aj, 114, 713 

  \bibitem[Aller (1984)]{aller84} Aller, L.~H.\ 1984, Physics of Thermal Gaseous Nebulae, Reidel, Dordrech
  
  \bibitem[Andrievsky et al. (2004)]{andri04} Andrievsky, S.~M., Luck, R.~E., Martin, P., \& L{\'e}pine, J.~R.~D.\ 2004, \aap, 413, 159 

  \bibitem[Ballero et al. (2007)]{bal07} Ballero, S.~K., Matteucci, F., Origlia, L., \& Rich, R.~M.\ 2007, \aap, 467, 123 
 
  \bibitem[Cahn et al. (1992)]{cahn92} Cahn, J.~H., Kaler, J.~B., \& Stanghellini, L.\ 1992, \aaps, 94, 399 

  \bibitem[Carigi et al. (2005)]{carigi05} Carigi, L., Peimbert,  M., Esteban, C., \& Garc{\'{\i}}a-Rojas, J.\ 2005, \apj, 623, 213 

  \bibitem[Cardelli et al. (1989)]{ccm89} Cardelli, J.~A., Clayton, G.~C., \& Mathis, J.~S.\ 1989, \apj, 345, 245 

  \bibitem[Chiappini et al. (2001)]{chiap01} Chiappini, C., Matteucci, F., \& Romano, D.\ 2001, \apj, 554, 1044 	

  \bibitem[Chiappini et al. (2009)]{chiap09} Chiappini, C., Gorny, S., Stasi\'nska, G., \& Barbuy, B.\ 2009, \aap, 494, 591

  \bibitem[Costa et al. (1996)]{costa96} Costa, R.~D.~D., Chiappini, C., Maciel, W.~J., \& de Freitas Pacheco, J.~A.\ 1996, \aaps, 116, 249 
  
  \bibitem[Costa et al. (2000)]{costa00} Costa, R.~D.~D., de Freitas Pacheco, J.~A., \& Idiart, T.~P.\ 2000, \aaps, 145, 467 

  \bibitem[Costa et al. (2005)]{costa05} Costa, R.~D.~D., Escudero, A.~V., \& Maciel, W.~J.\ 2005, in Planetary Nebulae as Astronomical Tools, ed. R. Szczerba, G. Stasi\'nska, S. K. G\'orny, AIP Conf. Proc., 804, 252 

  \bibitem[Costa \& Maciel (2006)]{costa06} Costa, R.~D.~D., \& Maciel, W.~J.\ 2006, in Planetary Nebulae in our Galaxy and Beyond, ed. M. J. Barlow, \& R. H. M\'endez (Cambridge: CUP), IAU Symp., 234, 243

  \bibitem[Costa et al. (2008)]{costa08} Costa, R.~D.~D., Maciel, W.~J., \& Escudero, A.~V.\ 2008, Baltic Astronomy, 17, 321

  \bibitem[Cuisinier et al. (1996)]{cuisin96} Cuisinier, F., Acker, A., \& Koeppen, J.\ 1996, \aap, 307, 215 

  \bibitem[Cuisinier et al. (2000)]{cuisin00} Cuisinier, F., Maciel, W.~J., K{\"o}ppen, J., Acker, A., \& Stenholm, B.\ 2000, \aap, 353, 543 

  \bibitem[Daflon \& Cunha (2004)]{daflon04} Daflon, S., \& Cunha, K.\ 2004, \apj, 617, 1115 

  \bibitem[Escudero \& Costa (2001)]{escu01} Escudero, A.~V., \& Costa, R.~D.~D.\ 2001, \aap, 380, 300 

  \bibitem[Escudero et al. (2004)]{escu04} Escudero, A.~V., Costa, R.~D.~D., \& Maciel, W.~J.\ 2004, \aap, 414, 211 

  \bibitem[Exter et al. (2004)]{exter04} Exter, K.~M., Barlow, M.~J., \& Walton, N.~A.\ 2004, \mnras, 349, 1291 

  \bibitem[de Freitas-Pacheco et al. (1993)]{fp93} de Freitas-Pacheco, J.~A., Barbuy, B., Costa, R.~D.~D., \& Idiart, T.~E.~P.\ 1993, \aap, 271, 429

  \bibitem[Fitzpatrick (1999)]{fitz99} Fitzpatrick, E.~L.\ 1999, \pasp, 111, 63 

  \bibitem[Girard et al. (2007)]{gir07} Girard, P., K{\"o}ppen, J., \& Acker, A.\ 2007, \aap, 463, 265 

  \bibitem[G{\'o}rny et al. (2004)]{gorny04} G{\'o}rny, S.~K., Stasi{\'n}ska, G., Escudero, A.~V., \& Costa, R.~D.~D.\ 2004, \aap, 427, 231 

  \bibitem[G{\'o}rny et al. (2009)]{gorny09} G{\'o}rny, S.~K., Chiappini, C., Stasinska, G., \& Cuisinier, F.\ 2009, 500, 1089

  \bibitem[Gutenkunst et al. (2008)]{guten08} Gutenkunst, S., Bernard-Salas, J., Pottasch, S.~R., Sloan, G.~C., \& Houck, J.~R.\ 2008, \apj, 680, 1206 

  \bibitem[Kingsburgh \& Barlow (1994)]{kings94} Kingsburgh, R.~L., \& Barlow, M.~J.\ 1994, \mnras, 271, 257 

  \bibitem[Kingdon \& Ferland (1995)]{king95} Kingdon, J., \& Ferland, G.~J.\ 1995, \apj, 442, 714

  \bibitem[Lattanzio \& Forestini (1999)]{lat99} Lattanzio, J., \& Forestini, M.\ 1999, in Asymptotic Giant Branch Stars, ed. T. Le Bertre, A. L\`ebre, C. Waelkens, (San Francisco: ASP), IAU Symp. 191, 31 
  
  \bibitem[Lecureur et al. (2007)]{lecu07} Lecureur, A., Hill, V., Zoccali, M., Barbuy, B., G{\'o}mez, A., Minniti, D., Ortolani, S., \& Renzini, A.\ 2007, \aap, 465, 799 
  
  \bibitem[Maciel et al. (2005)]{maciel05} Maciel, W.~J., Lago, L.~G., \& Costa, R.~D.~D.\ 2005, \aap, 433, 127 
  
  \bibitem[Maciel et al. (2006)]{maciel06} Maciel, W.~J., Lago, L.~G., \& Costa, R.~D.~D.\ 2006, \aap, 453, 587   

  \bibitem[Maciel \& Pottasch (1980)]{maciel80} Maciel, W.~J., \& Pottasch, S.~R.\ 1980, \aap, 88, 1 

  \bibitem[Maciel \& Quireza (1999)]{maciel99} Maciel, W.~J., \& Quireza, C.\ 1999, \aap, 345, 629 

  \bibitem[Mathis (1990)]{mathis90} Mathis, J.~S.\ 1990, \araa, 28, 37 

  \bibitem[Peimbert \& Serrano (1980)]{peimb80} Peimbert, M., \& Serrano, A.\ 1980, Revista Mexicana de Astronomia y Astrofisica, 5, 9 

  \bibitem[Pequignot et al. (1991)]{pequi91} Pequignot, D., Petitjean, P., \& Boisson, C.\ 1991, \aap, 251, 680 

  \bibitem[Perinotto (1991)]{peri91} Perinotto, M.\ 1991, \apjs, 76, 687 

  \bibitem[Perinotto et al. (2004)]{peri04} Perinotto, M., Morbidelli, L., \& Scatarzi, A.\ 2004, \mnras, 349, 793 

  \bibitem[Perinotto \& Morbidelli (2006)]{peri06} Perinotto, M., \& Morbidelli, L.\ 2006, \mnras, 372, 45 

  \bibitem[Ratag et al. (1992)]{ratag92} Ratag, M.~A., Pottasch, S.~R., Dennefeld, M., \& Menzies, J.~W.\ 1992, \aap, 255, 255  

  \bibitem[Ratag et al. (1997)]{ratag97} Ratag, M.~A., Pottasch, S.~R., Dennefeld, M., \& Menzies, J.\ 1997, \aaps, 126, 297  

  \bibitem[Rich (1988)]{rich88} Rich, R.~M.\ 1988, AJ, 95, 828 

  \bibitem[Rich \& Origlia (2005)]{rich05} Rich, R.~M., \& Origlia, L.\ 2005, \apj, 634, 1293

  \bibitem[Stanghellini et al. (2008)]{stangh08} Stanghellini, L., Shaw, R.~A., \& Villaver, E.\ 2008, \apj, 689, 194 

  \bibitem[Stasi{\'n}ska et al. (1998)]{stas98} Stasi{\'n}ska, G., Richer, M.~G., \& McCall, M.~L.\ 1998, \aap, 336, 667 
 
  \bibitem[Torres-Peimbert \& Peimbert (1977)]{tpp77} Torres-Peimbert, S., \& Peimbert, M.\ 1977, Revista Mexicana de Astronomia y Astrofisica, 2, 181 

  \bibitem[van Loon et al. (2003)]{vanloon03} van Loon, J.~T., Gilmore, G.~F., Omont, A., et al.\ 2003, \mnras, 338, 857 

  \bibitem[Wang \& Liu (2007)]{wang07} Wang, W., \& Liu, X.-W.\ 2007, \mnras, 381, 669 
  
  \bibitem[Weiland et al. (1994)]{wei94} Weiland, J. L., Arendt, R. G., Berriman, G. B., et al.\ 1994, \apjl, 425, L81

  \bibitem[Wesson et al. (2005)]{wesson05} Wesson, R., Liu, X.-W., \& Barlow, M.~J.\ 2005, \mnras, 362, 424 

  \bibitem[Zoccali et al. (2006)]{zoc06} Zoccali, M., Lecureur, A., Barbuy, B., et al.\ 2006, \aap, 457, L1
  
  \bibitem[Zoccali et al. (2003)]{zoc03} Zoccali, M., Renzini, A., Ortolani, S., et al.\ 2003, \aap, 399, 931 

\end{thebibliography}
\end{document}